\documentclass[10pt,twocolumn,letterpaper]{article}

\usepackage[arxiv]{iccv}      

\usepackage{url}
\usepackage{graphicx}
\usepackage{booktabs}
\usepackage{xcolor}
\usepackage{adjustbox}
\usepackage{makecell}
\usepackage{tikz}
\usepackage[most]{tcolorbox}
\usepackage{xcolor}
\usepackage{multirow}
\usepackage{enumitem}
\usepackage{caption}
\usepackage{subcaption}
\usepackage{amssymb}
\usepackage{diagbox}
\usepackage{verbatim}
\usepackage{wrapfig}
\usepackage{pifont}

\definecolor{lightgray}{gray}{0.9}
\definecolor{lightlightgray}{gray}{0.95}
\definecolor{WildStrawberry}{rgb}{1.0, 0.26, 0.64}
\definecolor{Air Force blue}{rgb}{0.36, 0.54, 0.66}
\definecolor{cvprblue}{rgb}{0.21,0.49,0.74}
\usepackage[pagebackref,breaklinks,colorlinks,allcolors=cvprblue]{hyperref}
\usepackage{siunitx}
\usepackage[misc]{ifsym}
\def\confName{ICCV}
\def\confYear{2025}

\title{Towards Understanding Graphical Perception in Large Multimodal Models}

\author{
$^2$Kai Zhang\thanks{Work Done during internship at Microsoft Research.} \thanks{\Letter~:~~zhang.13253@osu.edu, chenwang@microsoft.com}\quad
$^1$Jianwei Yang\quad
$^1$Jeevana Priya Inala\quad
$^1$Chandan Singh\quad
$^1$Jianfeng Gao\\
$^2$Yu Su\quad
$^1$Chenglong Wang\footnotemark[2]\\[2mm]
$^1$Microsoft Research~~~~~~
$^2$The Ohio State University
}

\begin{document}

\maketitle

\begin{abstract}
Despite the promising results of large multimodal models (LMMs) in complex vision-language tasks that require knowledge, reasoning, and perception abilities together, we surprisingly found that these models struggle with simple tasks on infographics that require perception only.
As existing benchmarks primarily focus on end tasks that require various abilities, they provide limited, fine-grained insights into the limitations of the models' perception abilities.
To address this gap, we leverage the theory of graphical perception, an approach used to study how humans decode visual information encoded on charts and graphs, to develop an evaluation framework for analyzing gaps in LMMs' perception abilities in charts. With automated task generation and response evaluation designs, our framework enables comprehensive and controlled testing of LMMs' graphical perception across diverse chart types, visual elements, and task types.
We apply our framework to evaluate and diagnose the perception capabilities of state-of-the-art LMMs at three granularity levels (chart, visual element, and pixel). Our findings underscore several critical limitations of current state-of-the-art LMMs, including GPT-4o: their inability to (1) generalize across chart types, (2) understand fundamental visual elements, and (3) cross reference values within a chart.
These insights provide guidance for future improvements in perception abilities of LMMs.
The evaluation framework and labeled data are publicly available.\footnote{\href{https://github.com/microsoft/lmm-graphical-perception}{https://github.com/microsoft/lmm-graphical-perception}}

\begin{figure}
  \centering
  \includegraphics[width=0.42\textwidth]{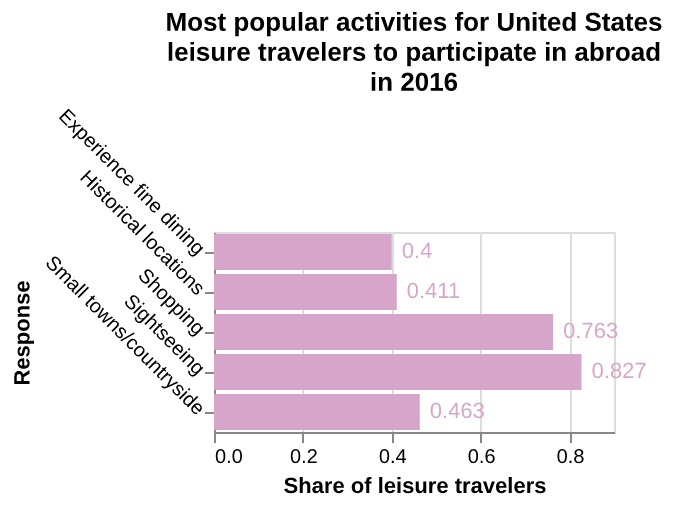} 
  \vspace{-0.5em}
  \caption{Given \textit{``Identify the activity with the highest share of leisure travelers.''}, GPT-4o responds \textit{``Small towns/countryside''} (10/10) whereas the correct answer is \textit{``Sightseeing.''}
  For \textit{``What's the share of travelers who go shopping?''}, the latest Claude-Sonnet-3.7 and Grok-3 reply \textit{``0.411''} (10/10), but the correct answer is \textit{``0.763.''}
Despite significant recent improvements in LMMs, their persistent perceptual limitations lead to mistakes on tasks that are trivial for humans.
  }
  \label{fig:teaser}
\vspace{-0.5em}
\end{figure}
\end{abstract}

\section{Introduction}

Large multimodal models (LMMs;~\cite{gpt-4o, deepmind_gemini_report}) have shown human-level results in a range of visual-language tasks~\citep{yue2023mmmu, lu2023mathvista}, including complex knowledge- and reasoning-intensive tasks over infographics in scientific documents.
However, we surprisingly found these LMMs may fail to solve some simple tasks on chart such as retrieving values or finding the extremum in given charts, where humans can easily achieve near-perfect results~\cite{Saket2019TaskEffectiveness}.
Figure~\ref{fig:teaser} shows an example.

Given that existing models have shown impressive object recognition and reasoning capabilities~\cite{yue2023mmmu}, we suspect existing LMMs' under-performing in these simple chart understanding tasks come from their limitations in \emph{graphical perception}~\cite{cleveland1984graphical}, the essential perceptual ability that human leverage to understand data-coded graphical elements in charts.
Existing benchmarks primarily focus on overall task performance, often combining perception, knowledge, and high-level reasoning into a single accuracy score.
This metric serves as an indirect proxy for measuring LMMs' perception  of charts—a high score may suggest good perception abilities, but poor performance makes it unclear whether the failure stems from perception errors, lack of knowledge, reasoning flaws, or a combination of these factors.
Additionally, models may not even need to perceive and reason about charts to answer questions: as reported in recent studies, models can generate correct answers without the visual input~\citep{yue2024mmmupro, chen2024mm-star}.
Finally, the way current models perceive charts—especially their understanding of fundamental visual elements—remains unexplored in existing work.
Thus, it is desirable to systematically study models' perception capabilities to understand factors limiting their performance in low-level chart perception tasks.

In this paper, we leverage \emph{graphical perception}~\citep{cleveland1984graphical}, a theory originally developed to study human interpretation of visual data in charts and graphs, to investigate models' perception capabilities.
For example, prior studies on human graphical perception show that, because humans can better perceive the length of lines than area sizes when comparing values, we can more efficiently and accurately read bar charts than pie charts to answer questions about calculating the differences between two values based on their visual representation in the charts.
This motivates us to evaluate models' fundamental graphical perception (e.g., perception of color, length, size) to explore the limitations of current LMMs.
To achieve this goal, we aim to test LMMs' performance on a range of chart perception tasks that involve reading and interpreting data based on its visual representation across a diverse set of chart types (e.g., bar, line, scatter, pie) and visual elements (e.g., color, length, size).
This approach could offer a direct and comprehensive evaluation of models' perceptual abilities, especially helping us understand in {what aspects the models fail to generalize}.

We introduce an evaluation framework specifically designed to assess the graphical perception abilities of state-of-the-art (SOTA) LMMs~\citep{gpt-4o, chen2024internvl2, Abdin2024phi3, meng2024chartassisstant}. Our framework includes an automated task generation and response evaluation pipeline that synthesizes a diverse set of chart perception tasks with different chart representations from a set of seed datasets, allowing us to scale up the evaluation with minimal human intervention.
With this framework, we evaluate SOTA LMMs in a coarse-to-fine manner, ranging from chart-type-level performance to the fundamental visual elements forming the charts, and to the pixel-level analysis that reveals how models perceive specific regions in the charts. Our goal is to understand where and how models fail to generalize in their perception of charts.
Our research questions and key findings are listed as follows. These findings provide fine-grained insights into the low-level visual abilities of LMMs from the perspective of graphical perception.

\smallskip
\textbf{RQ1: Can SOTA LMMs Generalize Across Diverse Chart Types?} LMMs exhibit significant performance fluctuations depending on the chart type and rely heavily on the explicit numerical annotations.
LMMs cannot generalize across different chart types, despite the simplicity and identical information presented.

\smallskip
\textbf{RQ2: Do LMMs Learn Generalizable and Compositional Visual Elements Beyond Chart Patterns?} LMMs perform relatively well only on charts with specific combinations of visual elements (e.g., length, size, position) but struggle to generalize to charts composed of similar visual elements.
This indicates that current LMMs do not develop compositional graphical perception by learning from common chart patterns, as their performance drops significantly when we compose new charts that require different combinations of perception dimensions.

\smallskip
\textbf{RQ3: Can We Explain LMMs' Limitations Based on Their Pixel-Level Perception Patterns?}
While models often successfully locate important regions required for solving simple tasks such as retrieving data point values, their referencing of these values is frequently imprecise, leading to only approximate outputs.
This imprecision accumulates in more complex tasks, such as ordering all the data points.

\section{Evaluation Framework}
\label{sec:experimental-setup}
\begin{figure*}[tb]
    \centering
    \includegraphics[width=\textwidth]{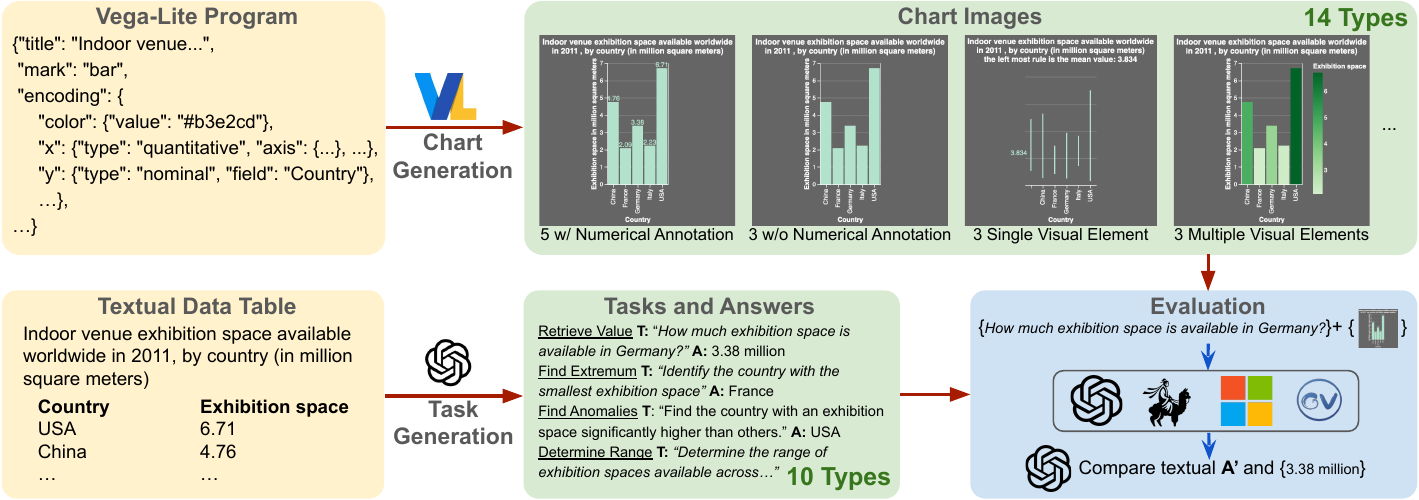}
    \caption{Framework of data synthesis and evaluation.
    With randomly sampled 1,000 datasets as seeds, we edit the Vega-Lite program to generate 14 types of charts and use GPT-4o with textual data tables to generate 10 types of tasks and corresponding answers, resulting in a total of 140,000 inputs for each model to be evaluated.
For evaluation, we consider the most representative models from four model categories and their responses are automatically evaluated by GPT-4o in text format.
    }
    \label{fig:workflow}
\end{figure*}

The major difference between prior evaluations~\citep{masry-etal-2022-chartqa, wang2024charxiv} and ours is that we don't aim to create a more challenging benchmark.
Instead, as shown in Figure~\ref{fig:workflow}, we propose a \textit{framework} that \textit{automatically} creates various charts with different visual elements to evaluate and diagnose the graphical perception abilities of current SOTA LMMs.

\subsection{Dataset Generation}

In our framework, we utilize the VisText~\citep{tang2023vistext} dataset as the primary data seed as it covers diverse domains including sports, news, etc.
It includes both textual data tables and exemplar Vega-Lite programs~\citep{Satyanarayan2017Vega} which can be used to generate diverse rasterized charts after light-weight editing.
For later experiments, we randomly sample 1,000 datasets from VisText (e.g., one dataset shown in Fig.~\ref{fig:workflow}), ensuring a wide variety of data types and relationships are represented.

The datasets we use include three major types of data attributes: (1) Nominal Attributes: Categorical variables that represent distinct labels without an inherent order (e.g., country, movie genres).
(2) Ordinal Attributes: Variables that have a meaningful order or ranking but no fixed intervals between values (e.g., movie ratings, years).
(3) Numerical Attributes: Continuous variables that allow for the calculation of differences and other mathematical operations (e.g., exhibition space in Figure~\ref{fig:workflow}).
To ensure simplicity, each dataset has at most two data dimensions (i.e., a nominal attribute paired with a numerical attribute or an ordinal attribute paired with a numerical attribute).
In addition, the number of data points is limited to 5.
These constraints allow us to evaluate the models' graphical perception capabilities without overwhelming them with complex visuals.

\subsection{Task Generation}
\label{sec:tasks}

Given seed datasets, we instruct GPT-4o to instantiate tasks based on the textual data table.
During model inference, only chart images and task texts are provided.
Following prior work~\citep{Saket2019TaskEffectiveness} on human graphical perception, we design 10 types of tasks for each dataset, as shown in Table~\ref{tab:task-details}.

\begin{table*}[tbh]
\centering
\small

\resizebox{\textwidth}{!}{

\begin{tabular}{p{0.28\textwidth}|p{0.69\textwidth}}
\toprule
\textbf{Task} & \textbf{Description} \\ \midrule
T1. Retrieve Value & Retrieve the value of a given attribute for a specific data point. \\
T2. Find Extremum & Identify the maximum or minimum value of a specified attribute. \\
T3. Find Anomalies & Detect anomalies in the dataset regarding a given relationship or expectation. \\
T4. Determine Range & Determine the range of values for a given attribute. \\
T5. Find Correlation & Identify any correlation between two data attributes. \\
T6. Compute Derived Value & Compute a derived value from a set of data points. \\
T7. Filter & Filter the data points based on specific conditions. \\
T8. Order & Order the data points according to a numerical attribute. \\
T9. Find Clusters & Find clusters of similar attribute values. \\
T10. Characterize Distribution & Characterize the distribution of a data attribute over a given set. \\ \bottomrule
\end{tabular}
}
\caption{All 10 task types, ranging from a single data point (T1) to an entire dataset (T10).}
\label{tab:task-details}
\vspace{-0.5em}
\end{table*}

These tasks are designed to cover a broad spectrum of graphical perception skills, ranging from a single data point (e.g., T1), to multiple data points (e.g., T4), and to an entire dataset (e.g., T10).
This design allows us to evaluate how well models handle increasing levels of task complexity.
Appendix~\ref{appendix:task-generation} shows the detailed task generation prompt.

\subsection{Evaluation Targets}

Given the extensive variety of LMMs with different vision and language backbones, a comprehensive evaluation and analysis of all models may not be feasible.
Therefore, we focus on four categories of models: proprietary, open-source, lightweight, and chart-specialized LMMs.
We select the most representative model from each category for detailed evaluation and study, based on the averaged results reported on prior chart-included benchmarks~\citep{wang2024charxiv, yue2023mmmu, lu2023mathvista}.\footnote{Though we focus on four models in the main paper, our framework can easily extend to other models, showing consistent conclusions (Table~\ref{tab:rebuttal_results}).}
\begin{itemize}[left=0pt]
    \item \textbf{GPT-4o}~\citep{gpt-4o}, one of the strongest proprietary general-purpose model, represents the SOTA in LMMs. Benchmarking GPT-4o allows us to evaluate the performance of the latest model in chart tasks, providing a reference point for comparison with other models in this domain.

    \item \textbf{InternVL2}~\citep{chen2024internvl2} is one of the best open-source general-purpose LMMs.
    It is built upon Llama3.1~\citep{llama3-1} and has a total of 76B parameters.
    Evaluating InternVL2 can show the gap between open-source models and GPT-4o. 

    \item \textbf{Phi-3.5-Vision}~\citep{Abdin2024phi3} is selected as a strong lightweight general-purpose LMM, with only 4.2B parameters.
    With Phi-3.5, we can evaluate whether models with smaller vision backbones can reach decent levels of perception.

    \item \textbf{ChartAssistant}~\citep{meng2024chartassisstant} is the best chart-specialist model.
    It is continually trained with the LLaVA-13B~\citep{liu2023llava1-5} on a massive amount of chart datasets, including the original VisText dataset.
    With this specialist model, we can measure the benefits of in-domain training in enhancing perception and generalization abilities.

\end{itemize}

\subsection{GPT-4o-Aided Evaluation}

We employ GPT-4o as an automated text evaluator, which is particularly useful when models being evaluated output varied answer formats, such as chain-of-thought reasoning format~\citep{wei2023CoT}, or when dealing with open-ended tasks.
GPT-4o evaluates responses by comparing the textual responses of models against the predefined answer, which is generated automatically by GPT-4o based on textual representation of the data and chart program that do not need visual perception.
The evaluation process is guided by a detailed rubric designed for different task types.
For example, in \textit{Retrieve Value} tasks, answers are considered accurate if they are within a 5\% margin of the correct value.
For order-based tasks, such as ranking items, the model must return the exact sequence expected, while other list-based tasks do not require specific ordering.
Evaluation outcomes are categorized into accurate, fair, skipped, inaccurate, and n/a.
Please refer to Appendix~\ref{appendix:evaluation-rubric} for more details.

To calibrate GPT-4o’s evaluation process, we use a 10-shot demonstration~\citep{Brown2020GPT3} that includes examples of textual data tables, tasks, reference answers, model responses, and expected evaluations.
This calibration helps ensure consistency and accuracy in evaluation.
We manually review GPT-4o’s and Claude’s evaluations of the same 200 questions for each of the four models.
Table~\ref{tab:evaluation-accuracy} shows GPT-4o can achieve 99.0\% evaluation accuracy on average without noticeable bias across different models.
This establishes a reliable foundation for our evaluation framework.

\section{RQ1: Can SOTA LMMs Generalize Across Diverse Chart Types?}
\begin{table*}[tbh]
\vspace{-0.5em}
\centering

\resizebox{0.96\textwidth}{!}{
\begin{tabular}{l@{\;}>{\centering\arraybackslash}p{0.12\textwidth}@{\;}>{\centering\arraybackslash}p{0.12\textwidth}@{\;}>{\centering\arraybackslash}p{0.12\textwidth}@{\;}>{\centering\arraybackslash}p{0.12\textwidth}@{\;}>{\centering\arraybackslash}p{0.12\textwidth}@{\;}>{\centering\arraybackslash}p{0.12\textwidth}@{\;}>{\centering\arraybackslash}p{0.12\textwidth}@{\;}>{\centering\arraybackslash}p{0.12\textwidth}@{\;}}

\toprule

\multirow{2}{*}{} &
\multicolumn{4}{c}{Single Element} &
\multicolumn{4}{c}{Multiple Elements} \\ \cmidrule(lr){2-5} \cmidrule(lr){6-9}

& Length ($\leftrightarrow$) &
Color (\tikz \fill[Air Force blue] (0,0) rectangle (0.2,0.2);) &
Size (\tikz \filldraw[fill=gray] (0,0) circle (3pt);) &
Position ($\star$)  &
\makecell{$\leftrightarrow$, $\star$} & 
\makecell{\tikz \filldraw[fill=gray] (0,0) circle (3pt);, $\star$} & 
\makecell{$\leftrightarrow$, \tikz \filldraw[fill=gray] (0,0) circle (3pt);, $\star$} &
\makecell{$\leftrightarrow$, \tikz \fill[Air Force blue] (0,0) rectangle (0.2,0.2);, \tikz \filldraw[fill=gray] (0,0) circle (3pt);, $\star$} \\  \cmidrule(lr){1-5} \cmidrule(lr){6-9}

\centering Toy Chart          &
\adjustbox{valign=m}{\includegraphics[width=0.075\textwidth]{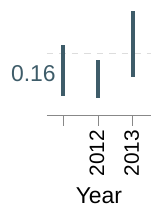}} & 
\adjustbox{valign=m}{\includegraphics[width=0.13\textwidth]{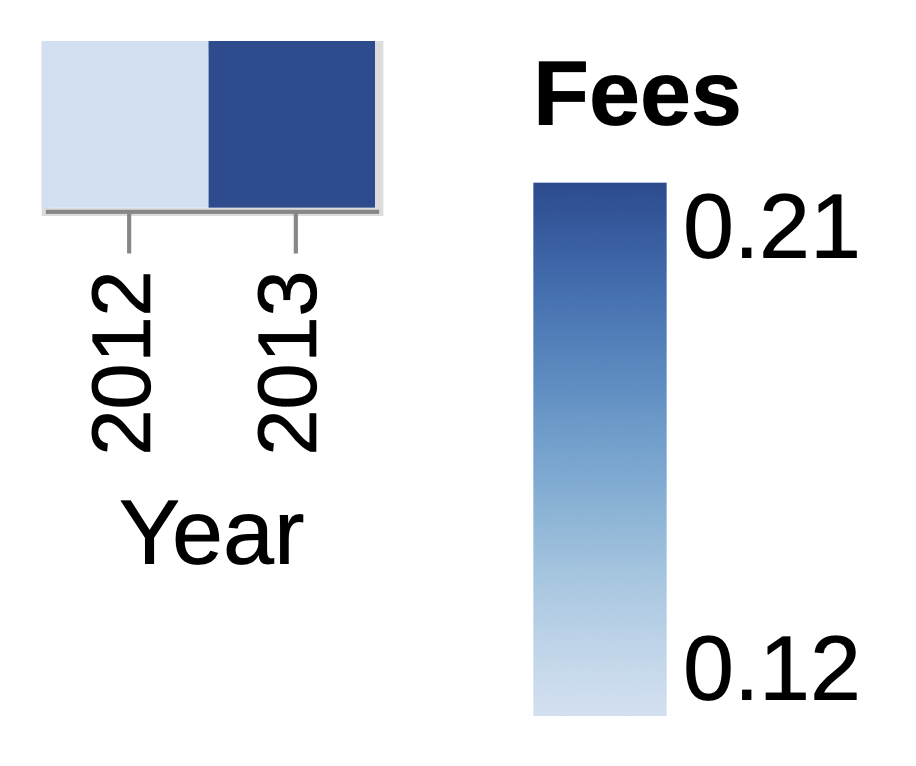}} & 
\adjustbox{valign=m}{\includegraphics[width=0.11\textwidth]{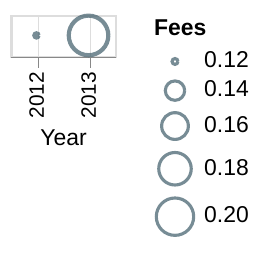}} &
\adjustbox{valign=m}{\includegraphics[width=0.09\textwidth]{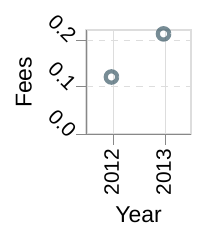}} &
\adjustbox{valign=m}{\includegraphics[width=0.09\textwidth]{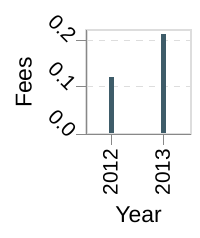}} & 
\adjustbox{valign=m}{\includegraphics[width=0.14\textwidth]{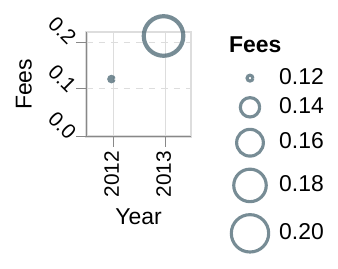}} & 
\adjustbox{valign=m}{\includegraphics[width=0.09\textwidth]{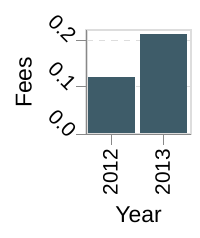}} & 
\adjustbox{valign=m}{\includegraphics[width=0.14\textwidth]{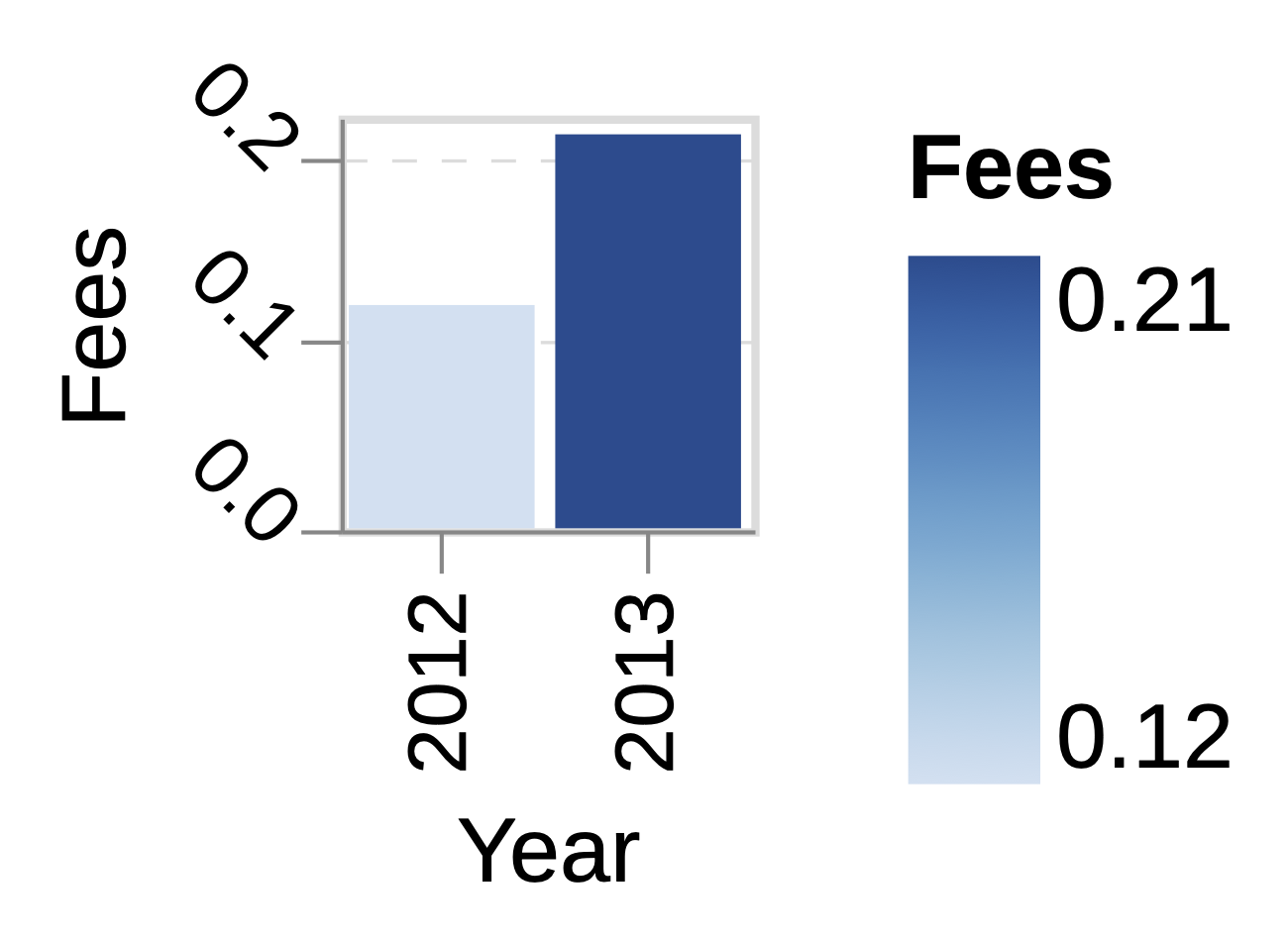}} \\  \cmidrule(lr){1-5} \cmidrule(lr){6-9}

GPT-4o            & 17.6                                                            & 21.1                                                           & 22.6                                                          & \textbf{41.2}   & 22.9 & 24.4 & \textbf{53.4} & 27.7                                                           \\

InternVL2      & 18.3                                                            & 20.7                                                           & 21.1                                                          & \textbf{33.4}  & 25.2 & 24.1 & \textbf{45.9} & 26.5  \\

Phi-3.5   & 17.8                                                            & 18.9                                                           & 19.6                                                          & \textbf{27.9}    & 21.2 & 20.9 & \textbf{32.4} & 22.7                                                          \\
ChartAssistant   & 12.3                                                            & 14.0                                                           & 13.6                                                          & \textbf{25.4}    & 19.2 & 17.4 & \textbf{33.9} & 18.5     \\ \bottomrule
\end{tabular}
}
\caption{Overall accuracy of models given charts rendering values with single or multiple visual elements. In multiple-element charts, a value is redundantly encoded through different elements. For example, the size, the position of top part, and the length of a bar are all proportional to the value.
In particular, we provide additional guidance to LMMs for reading uncommon charts.
}
\label{tab:single-and-multi-channel-performance}
\vspace{-0.5em}
\end{table*}

\label{sec:rq1}
\begin{tcolorbox}[title=Takeaways: No]

\begin{itemize}[left=0pt]
    \item Despite showing decent performance on specific chart types, LMMs struggle with variants of the same charts, showing limited generalization.
    \item LMMs heavily rely on explicit numerical annotations, performing significantly worse when annotations are removed.

\end{itemize}
\end{tcolorbox}

In this section, we analyze the performance of models at the chart-type level, where we compare models' performance in solving tasks with data represented in different chart types (line, bar, scatter, with and without explicit numerical annotations).
Despite the data being presented differently across charts, the models are expected to achieve similar performances due to the simplicity of the charts, similar to human performance (as shown in Table~\ref{tab:rebuttal_results}).
Figure~\ref{fig:appendix-chart-w-and-wo-anno-cases} shows chart examples used in this section.

\subsection{Analysis on Charts w/ Numerical Annotations}

\begin{figure}[htb]
    \centering
    \includegraphics[width=0.96\linewidth]{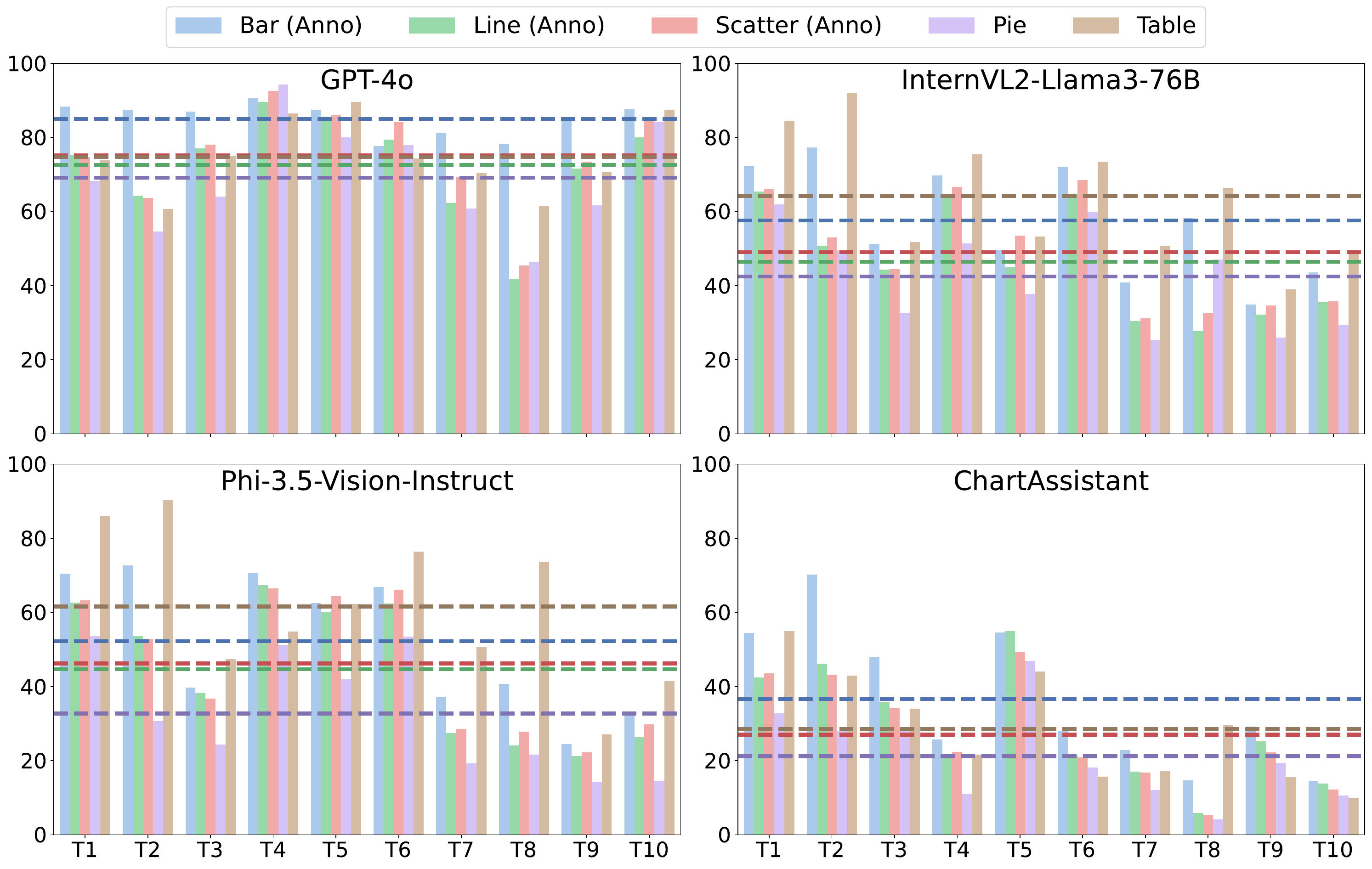}
    \caption{Accuracy of models on different types of charts with numerical annotations given the same 10 types of tasks.
    The dotted line refers to the average performance by chart type and color refers to the given chart type, and T-$i$ indicates the $i$-th task detailed in Section~\ref{sec:tasks}.}
\vspace{-0.5em}
    \label{fig:chart_type_eval_w_data_annotated}
\end{figure}

Figure~\ref{fig:chart_type_eval_w_data_annotated} presents the performance of models on different types of charts with numerical annotations given the same tasks, from which we can make the following observations:

(1) A significant performance gap exists for the same model when interpreting different types of charts containing the same information, suggesting that these models lack generalization across chart types.
For example, GPT-4o consistently performs the best overall, yet shows a clear preference for Bar (Anno) over Pie charts (85.0\% vs.\ 69.1\%).
This indicates that while GPT-4o excels in understanding some chart types, it still relies on specific visual structures to achieve its highest performance.
Meanwhile, InternVL2 and Phi-3.5, the open-source general-purpose models, perform best when presented with table images.
This observation suggests that these models might be specifically optimized for structured data.
However, the performance gaps of these models between different chart types are even larger, with up to 21.7\% for InternVL2 and 28.9\% for Phi-3.5, highlighting a stronger dependency on specific visual structures.

(2) Despite being trained on an extensive range of chart-related datasets and tested on simplified versions of its training data, ChartAssistant underperforms general-purpose models and struggles to generalize effectively across different chart types.
Its relatively better performance on Bar (Anno) can be attributed to the fact that bar charts make up 44.3\% of its training data.
This raises concerns about the effectiveness of chart-specific training for generalization.

(3) Models demonstrate significant performance variations across the ten tasks (T1 to T10), and task complexity amplifies the inconsistencies across chart types.
For example, GPT-4o performs relatively consistently across all tasks when given Bar (Anno) charts.
However, when interpreting Pie charts, its performance varies dramatically, with a gap of up to 48\% between simpler tasks like \textit{Determine Range} (T4, 93\%) and more complex tasks like \textit{Order} (T8, 46.3\%).
A similar trend can also be observed in the other two general-purpose open-source models.

These observations highlight the importance of improving graphical perception across a broader range of chart types to enhance LMMs' generalization in real-world applications where diverse charts appear in various forms.
Please refer to Appendix~\ref{appendix:full-results} for more detailed results.

\subsection{Analysis on Charts w/o Numerical Annotations}
\begin{table}[htb]
\centering

\resizebox{0.99\linewidth}{!}{
\begin{tabular}{lccc|ccc|ccc}
\toprule
               & \multicolumn{3}{c}{\textbf{Bar}} & \multicolumn{3}{c}{\textbf{Line}}                   & \multicolumn{3}{c}{\textbf{Scatter}} \\
               & w/ Anno.  & w/o Anno. & $\Delta$ & w/ Anno. & \multicolumn{1}{c}{w/o Anno.} & $\Delta$ & w/ Anno.   & w/o Anno.   & $\Delta$  \\ \midrule
GPT-4o        & 85.0      & 53.4      & \textcolor{WildStrawberry}{-31.6}    & 72.6     & 42.8                          & \textcolor{WildStrawberry}{-29.8}    & 75.1       & 41.2        & \textcolor{WildStrawberry}{-33.9}     \\ \midrule
InternVL2    & 57.6      & 45.9      & \textcolor{WildStrawberry}{-11.7}    & 46.4     & 33.0                          & \textcolor{WildStrawberry}{-13.4}    & 49.0       & 33.4        & \textcolor{WildStrawberry}{-15.6}     \\
Phi-3.5 & 52.2      & 32.4      & \textcolor{WildStrawberry}{-19.8}    & 44.7     & 26.0                          & \textcolor{WildStrawberry}{-18.7}    & 46.2       & 27.9        & \textcolor{WildStrawberry}{-18.3}     \\
ChartAssistant & 36.6      & 33.9      & \textcolor{WildStrawberry}{-2.7}     & 28.5     & 25.9                          & \textcolor{WildStrawberry}{-2.6 }    & 27.0       & 25.4        & \textcolor{WildStrawberry}{-1.6}      \\ \bottomrule
\end{tabular}
}
\caption{Overall accuracy of models given the charts with and without explicit numerical annotations.
Exemplar charts are shown in Figure~\ref{fig:workflow} (Chart Images) and Figure~\ref{fig:appendix-chart-w-and-wo-anno-cases}.
}
\vspace{-0.5em}
\label{tab:chart-type-eval-performance-changes}

\end{table}

Table~\ref{tab:chart-type-eval-performance-changes} shows the performance differences of models when transitioning from charts with numerical annotations (w/ Anno.) to charts without annotations (w/o Anno.).
GPT-4o shows the most significant drop in performance across all chart types when numerical annotations are removed, with an average performance decrease of 31.8\% across the three chart types.
This indicates that GPT-4o, despite being a leading model, still struggles to accurately perceive charts without the aid of numerical annotations.
Similarly, Phi-3.5 and InternVL2 exhibit substantial performance declines.

Additionally, Phi-3.5 shows a greater decline in performance compared to InternVL2 (e.g., -18.9\% vs. -13.6\% on average), showing that lightweight LMMs may have weaker generalization abilities than larger models when faced with charts lacking explicit numerical cues.
These observations show the importance of developing LMMs that are less reliant on numerical annotations as many complex charts in real-world scenarios do not include such annotations. See Appendix~\ref{appendix:full-results} for detailed results.

\section{RQ2: Do LMMs Learn Generalizable Visual Elements Beyond Chart Patterns?}
\begin{table*}[ht]
    \small
    \centering

\resizebox{\textwidth}{!}{
    \begin{tabular}{c|c}
        InternVL2 & Phi-3.5 \\[-8pt] \\  
        
        \begin{minipage}{0.235\linewidth}
            \centering
            
            \begin{tabular}{lcc}
                \toprule
                & $\blacksquare$ & $\square$ \\ \midrule
                Correct & 52 & 2 \\
                Incorrect & \textcolor{red}{34} & 12 \\ \bottomrule
            \end{tabular}
            \subcaption{Bar}
        \end{minipage}
        \hfill
        \begin{minipage}{0.235\linewidth}
            \centering
            
            \begin{tabular}{lcc}
                \toprule
                & $\blacksquare$ & $\square$ \\ \midrule
                Correct & 79 & 2 \\
                Incorrect & \textcolor{red}{13} & 6 \\ \bottomrule
            \end{tabular}
            \subcaption{Bar (Anno)}
        \end{minipage}
        \hfill
        & 
        \begin{minipage}{0.235\linewidth}
            \centering
            
            \begin{tabular}{lcc}
                \toprule
                & $\blacksquare$ & $\square$ \\ \midrule
                Correct & 29 & 6 \\
                Incorrect & \textcolor{red}{51} & 14 \\ \bottomrule
            \end{tabular}
            \subcaption{Bar}
        \end{minipage}
        \hfill
        \begin{minipage}{0.235\linewidth}
            \centering
            
            \begin{tabular}{lcc}
                \toprule
                & $\blacksquare$ & $\square$ \\ \midrule
                Correct & 72 & 1 \\
                Incorrect & \textcolor{red}{23} & 4 \\ \bottomrule
            \end{tabular}
            \subcaption{Bar (Anno)}
        \end{minipage}
    \end{tabular}
}
\vspace{-1em}
\caption{
Correctness at retrieving values in the \textit{Retrieve Value} task (table rows) depends on whether an LMM correctly identifies important chart regions (table columns).
    Identifying important regions is measured by whether the groundtruth labeled regions are covered ($\blacksquare$) or not covered ($\square$) by the LMM's feature importance map for value retrieval.
    Important regions are successfully identified more often for Bar (Anno) charts (b \& d).
    Sometimes, important regions are successfully identified but the model fails to retrieve the correct value (\textcolor{red}{red}).
    }
\label{tab:probing-matrix}
\vspace{-1.5em}
\end{table*}

\label{sec:rq2}
\begin{tcolorbox}[title=Takeaways: No. Superficial Chart Patterns Only]

\begin{itemize}[left=0pt]

    \item LMMs achieve relatively decent performance only when given specific combinations of visual elements but struggle even when generalizing to very similar charts, showing their lack of robust understanding of fundamental visual elements.
\end{itemize}
\end{tcolorbox}

Visual elements~\citep{bertin1967semiology, cleveland1984graphical, Munzner2014Vis2Analysis} are the core building blocks of data visualization, defining how quantitative values in charts are visualized.
Following prior work on human, we use four fundamental visual elements that are widely used to represent data values in charts: the position of a point (e.g., the top part of a bar), the length of a rule (e.g., bars or lines), the size of a region (e.g., the area of a bar), and the saturation of a color.
By systematically analyzing models' results on charts composed of these elements, we aim to assess how each element—or a combination thereof—impacts model perception, identifying which visual elements are most effective or challenging for current models. 
Particularly, as some of the generated charts may not be common, we provide guidelines on how to interpret these charts for LMMs.

Table~\ref{tab:single-and-multi-channel-performance} presents the performance of models when interpreting charts rendered with single or multiple basic visual elements.
We make three observations:

(1) LMMs suffer from basic visual element understanding.
Across the board, models show relatively poor performance when interpreting charts that rely on a single visual element, such as length, color, or size.
For example, GPT-4o achieves only 17.6\% accuracy on charts using length alone, despite its otherwise strong performance.
This indicates a fundamental challenge for LMMs in extracting quantitative values from basic visual elements, potentially limiting their abilities when comprehending complex charts where such basic visual elements are used.

(2) Surprisingly, the addition of redundant visual elements often hurt model performance.
For example, while using position only results in decent performance (e.g., GPT-4o scores with 41.2\% accuracy), rendering values via size at the same time (\tikz \filldraw[fill=gray] (0,0) circle (3pt);, $\star$) hurts the performance dramatically across all models.
Although the size can be more straightforward than position for tasks like ordering, LMMs clearly fail to leverage the advantages of various visual elements in most of the times.
This suggests that the presence of multiple visual elements may overwhelm the models' capacity to prioritize relevant visual cues, leading to confusion and misinterpretation of the data.

(3) LMMs often fail to generalize effectively across charts that use similar visual elements.
For instance, models show strong performance on bar charts that combine position, length, and size ($\leftrightarrow$, \tikz \filldraw[fill=gray] (0,0) circle (3pt);, $\star$), but struggle with similar charts that only use position and length ($\leftrightarrow$, $\star$).
This suggests that models excel only with specific combinations of visual elements and lack the robustness needed to transfer this understanding to slightly altered visualizations.

Overall, these results demonstrate that current LMMs merely follow specific and superficial perception patterns for common charts such as scatter ($\star$) and bar (\(\leftrightarrow\), \tikz \filldraw[fill=gray] (0,0) circle (3pt);, \(\star\)), while struggling to generalize beyond these familiar chart patterns.
This highlights the necessity of improving models' understanding of fundamental visual elements beyond specific chart types, leading to better generalization and perception.
Please refer to Appendix~\ref{appendix:full-results} for detailed results.

\section{RQ3: Explanation of Limitations on Graphical Perception Patterns}

\begin{tcolorbox}[title=Takeaways: Imprecise Value Referencing]
\begin{itemize}[left=0pt]
    \item LMMs often correctly localize the important regions in the bar charts for value retrieval (e.g., data points, axes), but fail to accurately cross reference the specific values, especially in charts without explicit number annotations.
\end{itemize}
\end{tcolorbox}

\begin{figure*}[tb]
    \centering
    \begin{subfigure}[t]{0.31\textwidth}
        \centering
        \includegraphics[height=4.0cm]{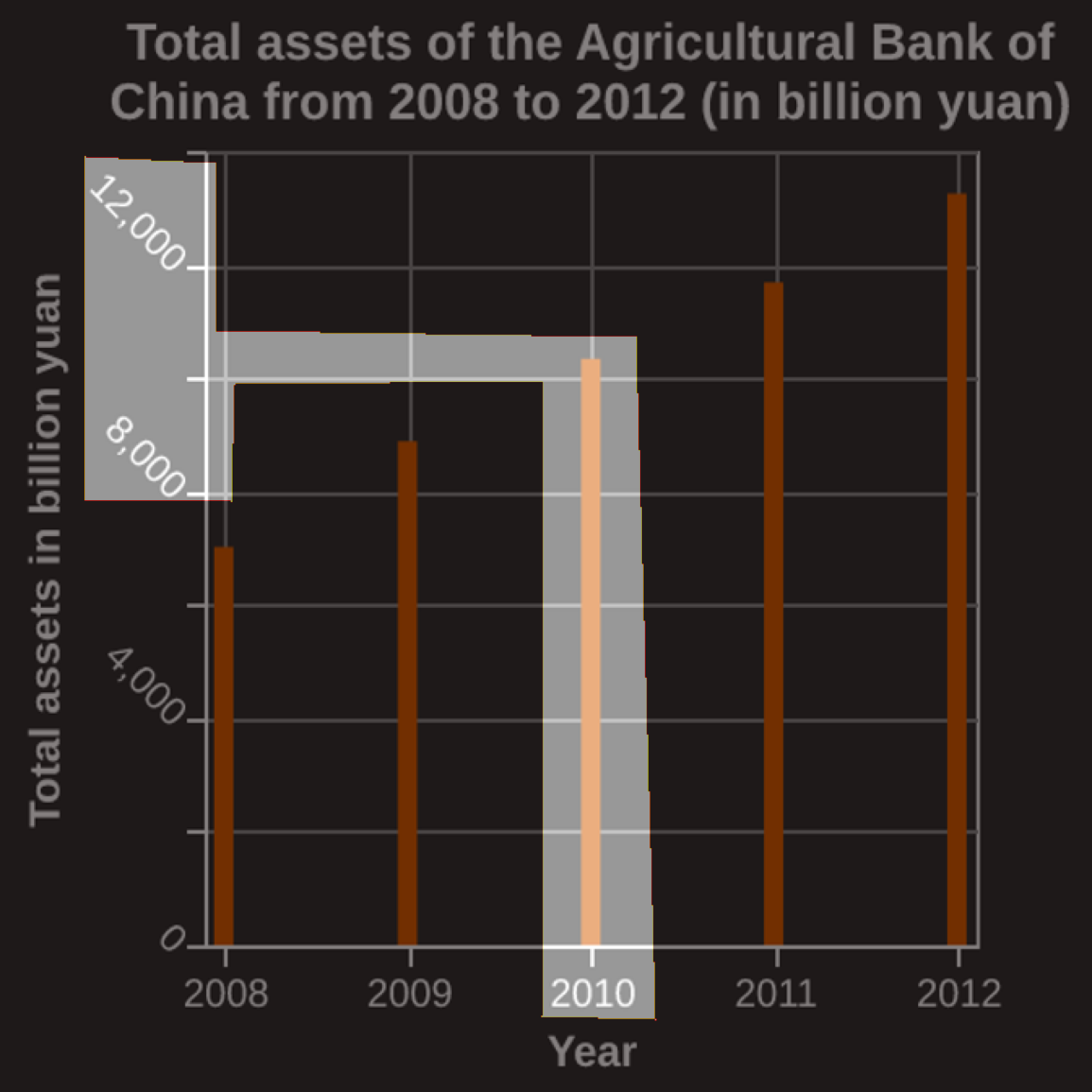}
        \caption{\makecell[c]{Labeled Regions\\Bar}}
        \label{fig:486-annotated-region-of-bar}
    \end{subfigure}
    \hfill
    \begin{subfigure}[t]{0.31\textwidth}
        \centering
        \includegraphics[height=4.0cm]{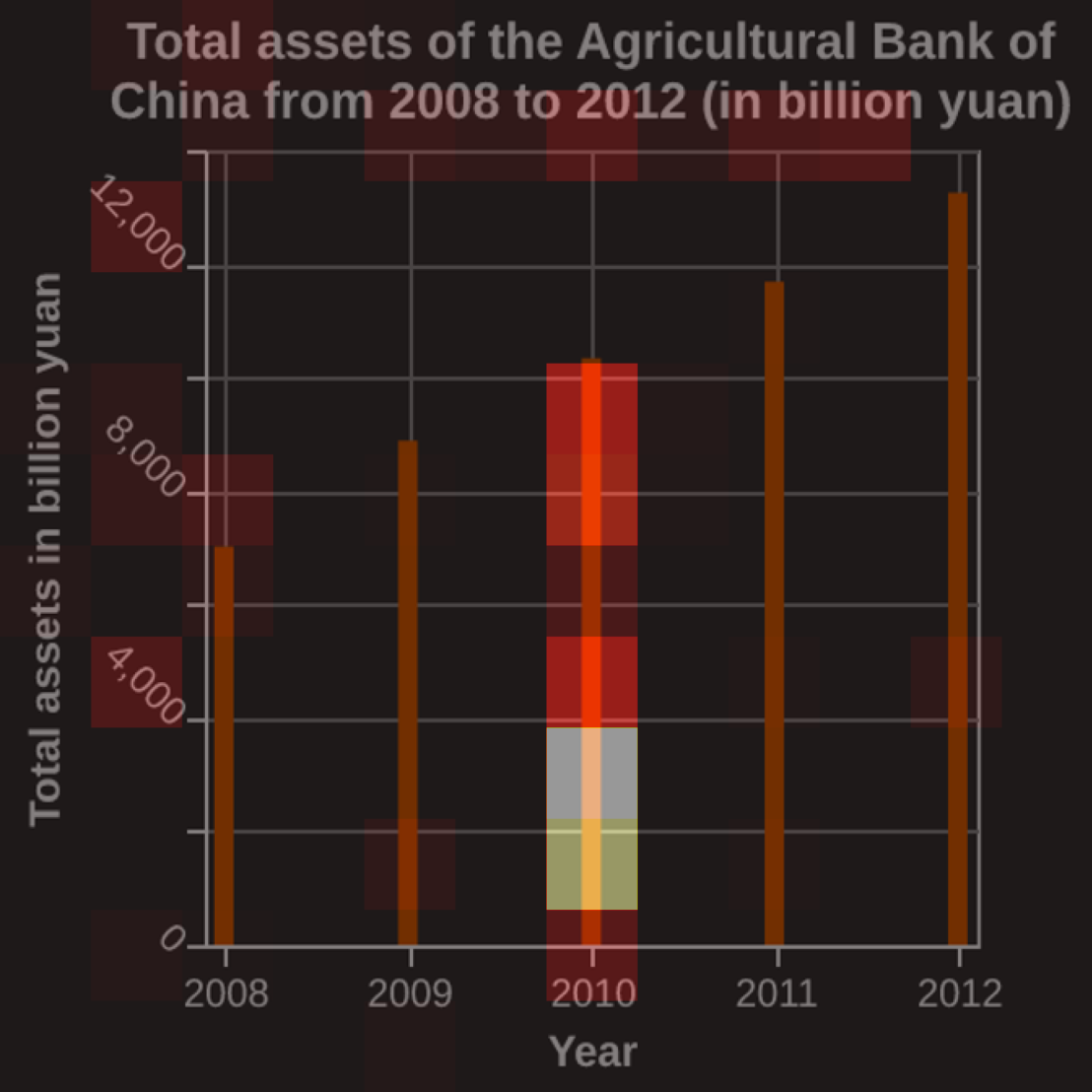}
        \caption{\makecell[c]{Heatmap-InternVL2\\Response: \textit{``9,000"} ($\blacksquare$)}}
        \label{fig:internvl-bar-covered-incorrect}
    \end{subfigure}
    \hfill
    \begin{subfigure}[t]{0.34\textwidth}
        \centering
        \includegraphics[height=4.0cm]{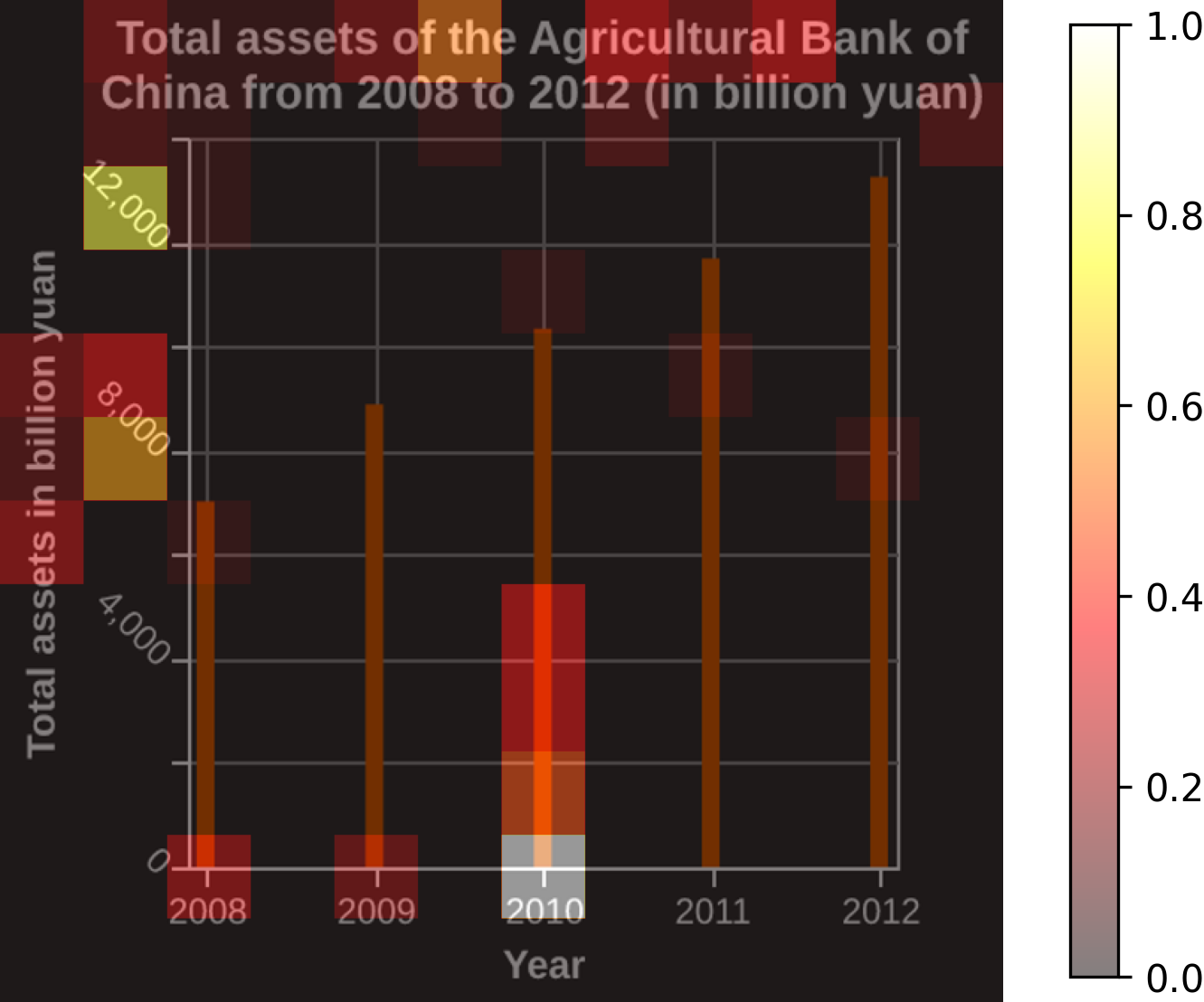}
        \caption{\makecell[c]{Heatmap-Phi-3.5\\Response: \textit{``9,000"}  ($\blacksquare$)}}
        \label{fig:phi35-bar-covered-incorrect}
    \end{subfigure}


    \begin{subfigure}[t]{0.31\textwidth}
        \centering
        \includegraphics[height=4.0cm]{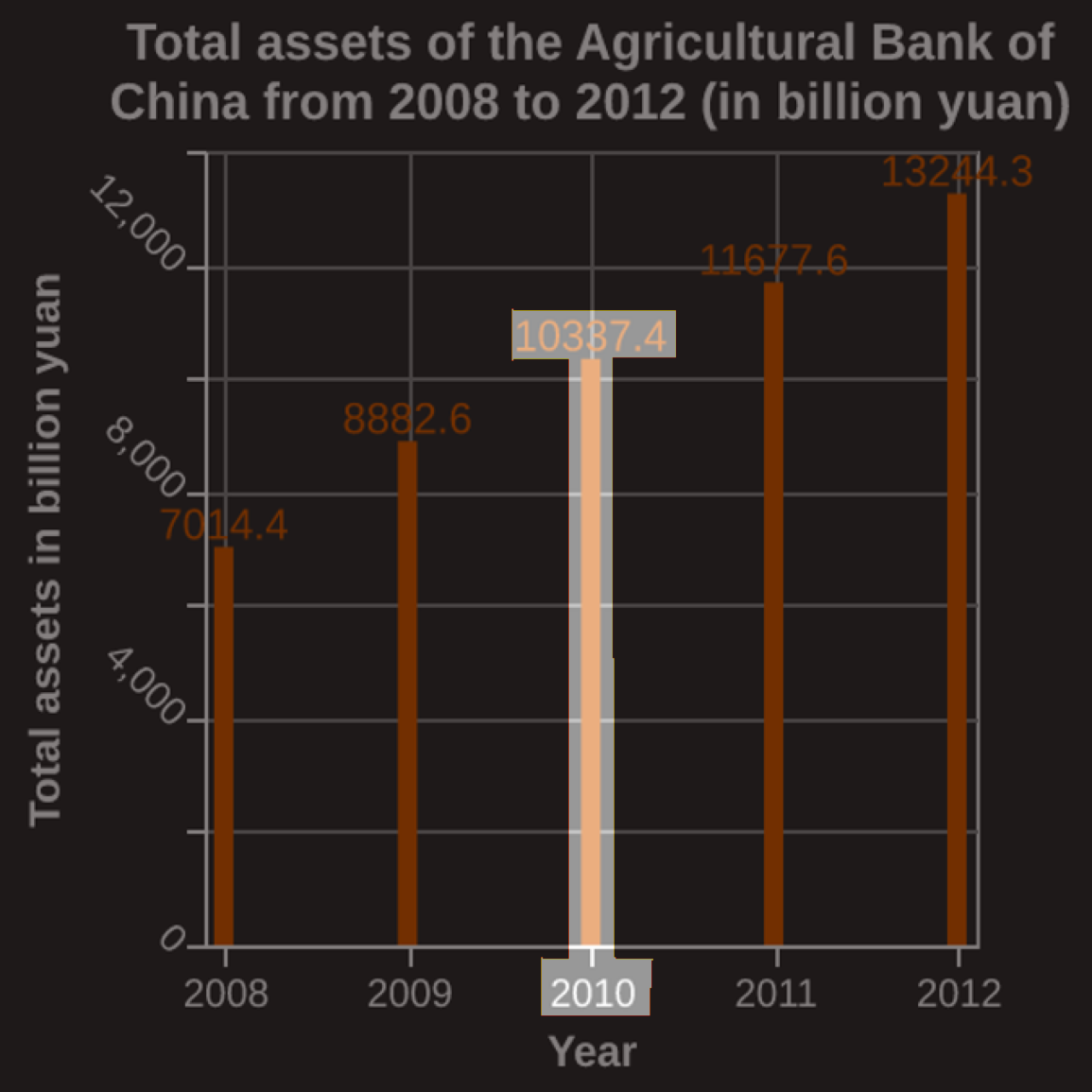}
        \caption{\makecell[c]{Labeled Regions\\Bar (Anno)}}
        \label{fig:486-annotated-region-of-bar-anno}
    \end{subfigure}
    \hfill
    \begin{subfigure}[t]{0.31\textwidth}
        \centering
        \includegraphics[height=4.0cm]{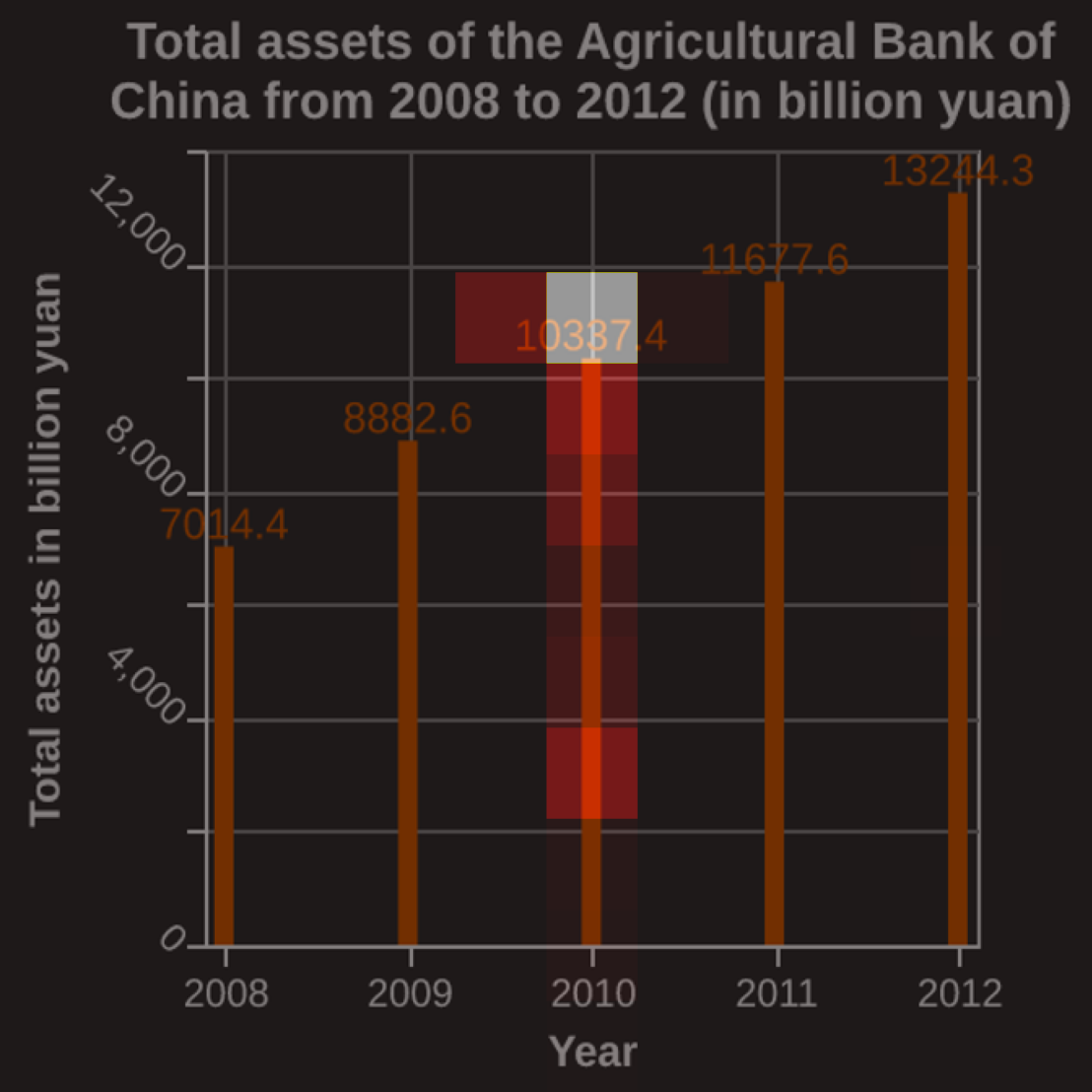}
        \caption{\makecell[c]{Heatmap-InternVL2\\Response: \textit{``10337.4"} ($\blacksquare$)}}
        \label{fig:internvl-bar-anno-covered-correct}
    \end{subfigure}
    \hfill
    \begin{subfigure}[t]{0.34\textwidth}
        \centering
        \includegraphics[height=4.0cm]{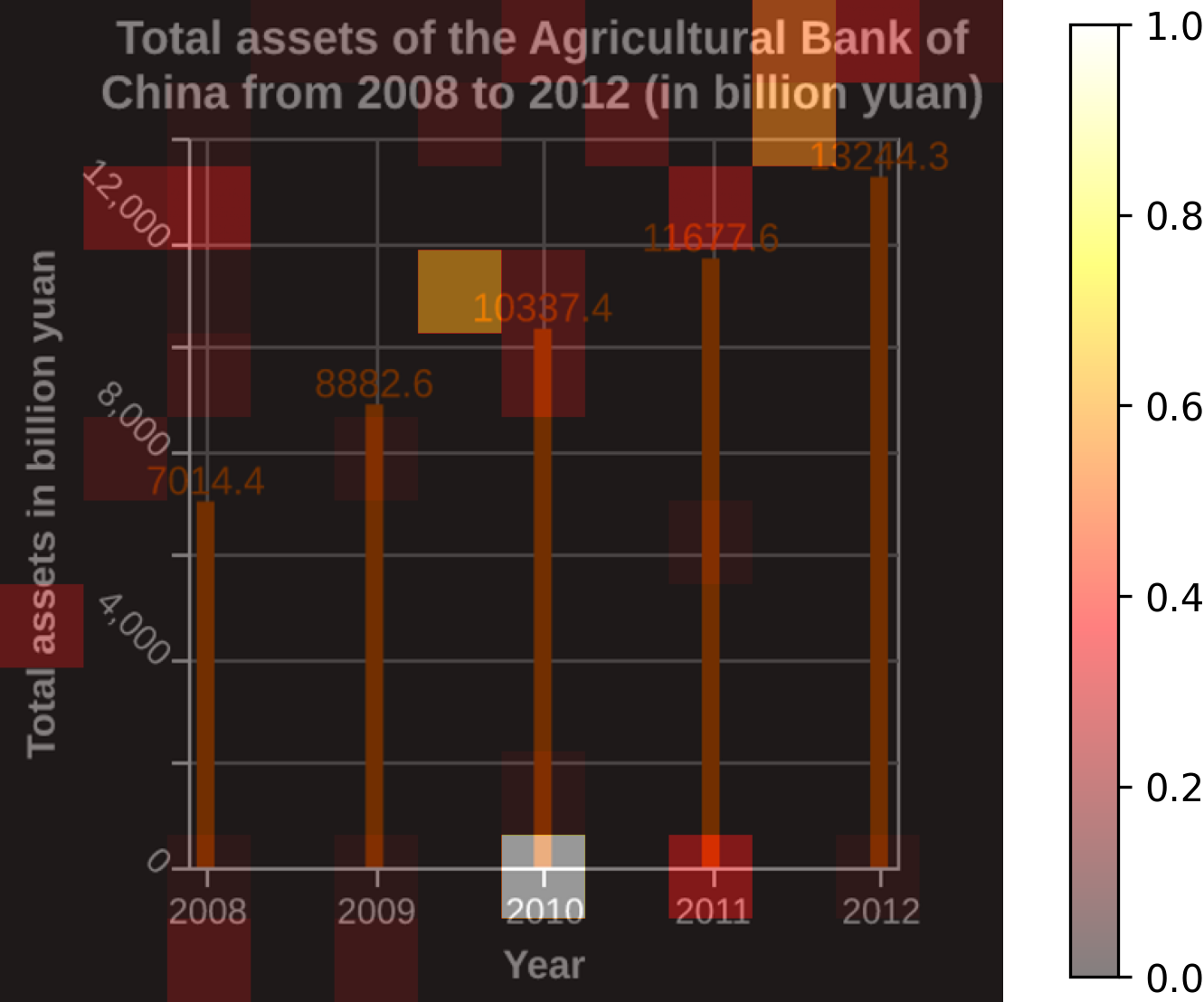}
        \caption{\makecell[c]{Heatmap-Phi-3.5\\Response: \textit{``10337.4"} ($\blacksquare$)}}
        \label{fig:phi35-bar-anno-covered-correct}
    \end{subfigure}
    \vspace{-1em}
    \caption{Examples of labeled regions and importance heatmaps for two models on Bar and Bar (Anno) charts.
    Given \textit{``What is the value of total assets in billion yuan for the year 2010?''}, both models successfully locate most labeled important regions on both the Bar (Anno) and Bar charts but fail to reference the correct y-axis values on the Bar chart.
    The correct answer is \textit{``10337.4.''}
    }
    \label{fig:case-study-covered-correct-and-wrong}
\vspace{-0.5em}
\end{figure*}

To understand the perception mechanisms of LMMs, we conduct a pixel-level analysis to examine which specific regions of the charts that models attend to when generating responses. 
This analysis aims to test whether LMMs correctly attend to important regions and cross reference the values of the chart for the most basic  \textit{Retrieve Value} task.

\subsection{Methodology}
Faithfully interpreting transformer-based models remains an open problem~\citep{bereska2024mechanisticInter, singh2024rethinkinginterpretability}, particularly in the context of newly emerging LMM capabilities.
In our analysis, we seek to use techniques that are model-agnostic, i.e., they can be applied to a black-box model without access to its weights or activations.\footnote{We also qualitatively explored some gradient- and attention-based interpretation methods~\citep{Selvaraju2017GradCam,wiegreffe2019attention}, but found that they were extremely sensitive to hyperparameters in the interpretation methods so we omit this analysis.
}
The most popular model-agnostic interpretation techniques generally require many calls to the model with different corruptions~\citep{ribeiro2016should,lundberg2017unified}, which are computationally expensive for large LMMs like GPT-4o.
We use a simplified, more efficient version of these methods that occludes different regions of the input image one at a time and measures the model's response.
Specifically, we manually select 100 pairs of Bar and Bar (Anno) charts and label the important regions for the \textit{Retrieve Value} task.
Each image is divided into 144 non-overlapping regions and we corrupt each region one at a time by changing its pixels to the background color.
We then calculate the difference in generated token logits between the intact chart and the corrupted version for each model.
We use the normalized logit difference as a measure of the feature importance of the region for the generated tokens.
We aggregate these region-level feature importances into a heatmap and measure whether the high-importance regions cover most of the groundtruth labeled regions (Intersection over Union $\geq$ 50\%) to determine whether models use these important regions.

\subsection{Results}
Table~\ref{tab:probing-matrix} shows that both InternVL2 and Phi-3.5 are quite effective at localizing important regions when given Bar (Anno) charts.
As long as the models can identify regions with the correct numbers, they generally generate correct responses, proving their reliance on explicit number annotations for accurate value retrieval shown in Section~\ref{sec:rq1}.
When annotations are removed (Bar), both models often still correctly locate the important regions but struggle to precisely refer to the values from the value-axis.
Figure~\ref{fig:case-study-covered-correct-and-wrong} illustrates this behavior.
In the Bar (Anno) chart (Figures~\ref{fig:486-annotated-region-of-bar-anno}-\ref{fig:phi35-bar-anno-covered-correct}), both models accurately identify important regions and generate correct responses.
In contrast, on Bar charts (Figures~\ref{fig:486-annotated-region-of-bar}-\ref{fig:phi35-bar-covered-incorrect}), although they focus on the right areas, their responses are far from correct.
In addition, Phi-3.5 tends to be more easily influenced by non-important regions compared to InternVL2, showing that lightweight models may be more sensitive to visual information irrelevant to the given task, leading to less favorable results shown in Figure~\ref{fig:chart_type_eval_w_data_annotated}.

\begin{figure*}[tb]
    \centering
    \begin{subfigure}[t]{0.46\textwidth}
        \centering
        \includegraphics[height=5cm]{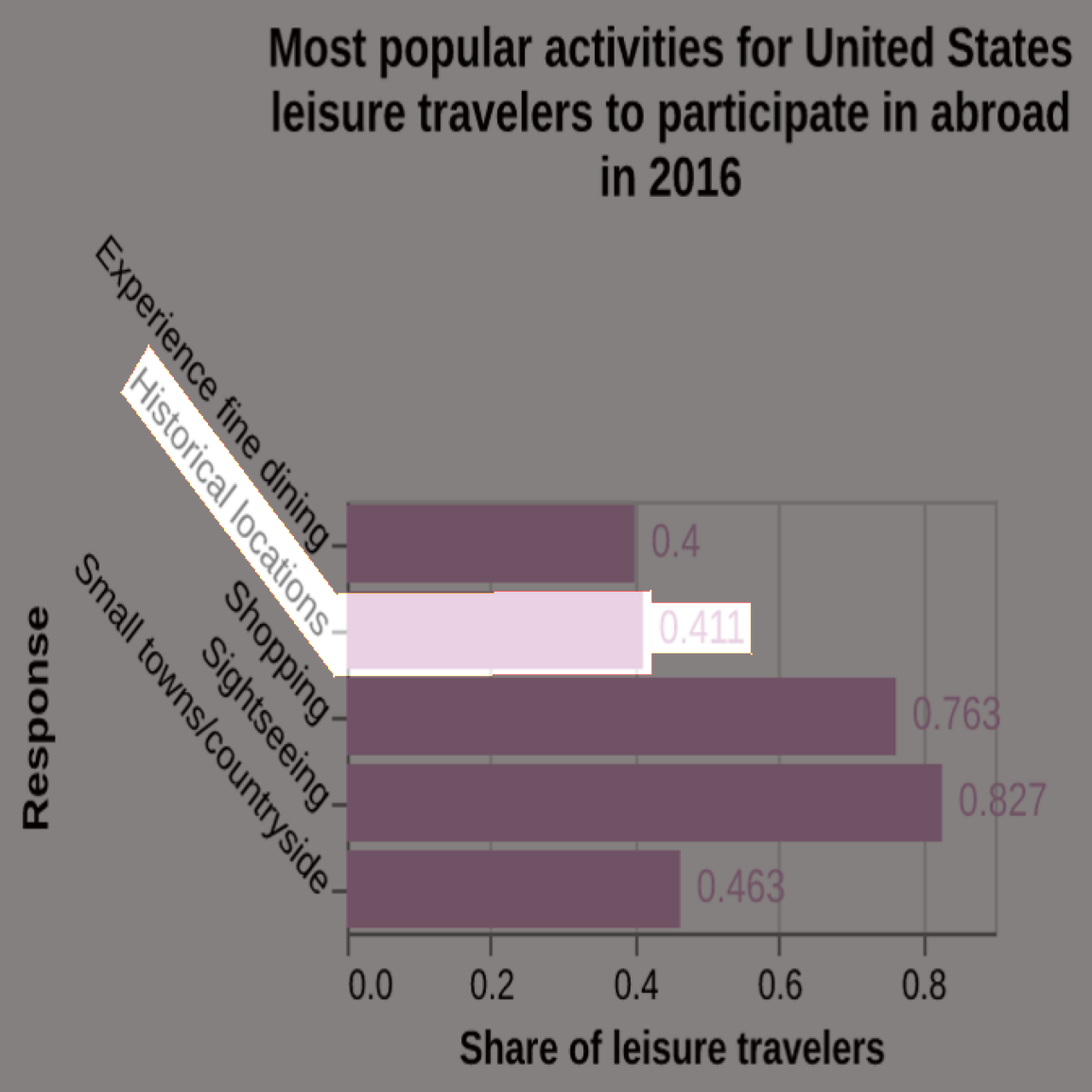}
        \caption{Labeled Regions-Bar (Anno)}
        \label{fig:1614-annotated-region-of-bar-anno}
    \end{subfigure}
    \hfill
    \begin{subfigure}[t]{0.46\textwidth}
        \centering
        \includegraphics[height=5cm]{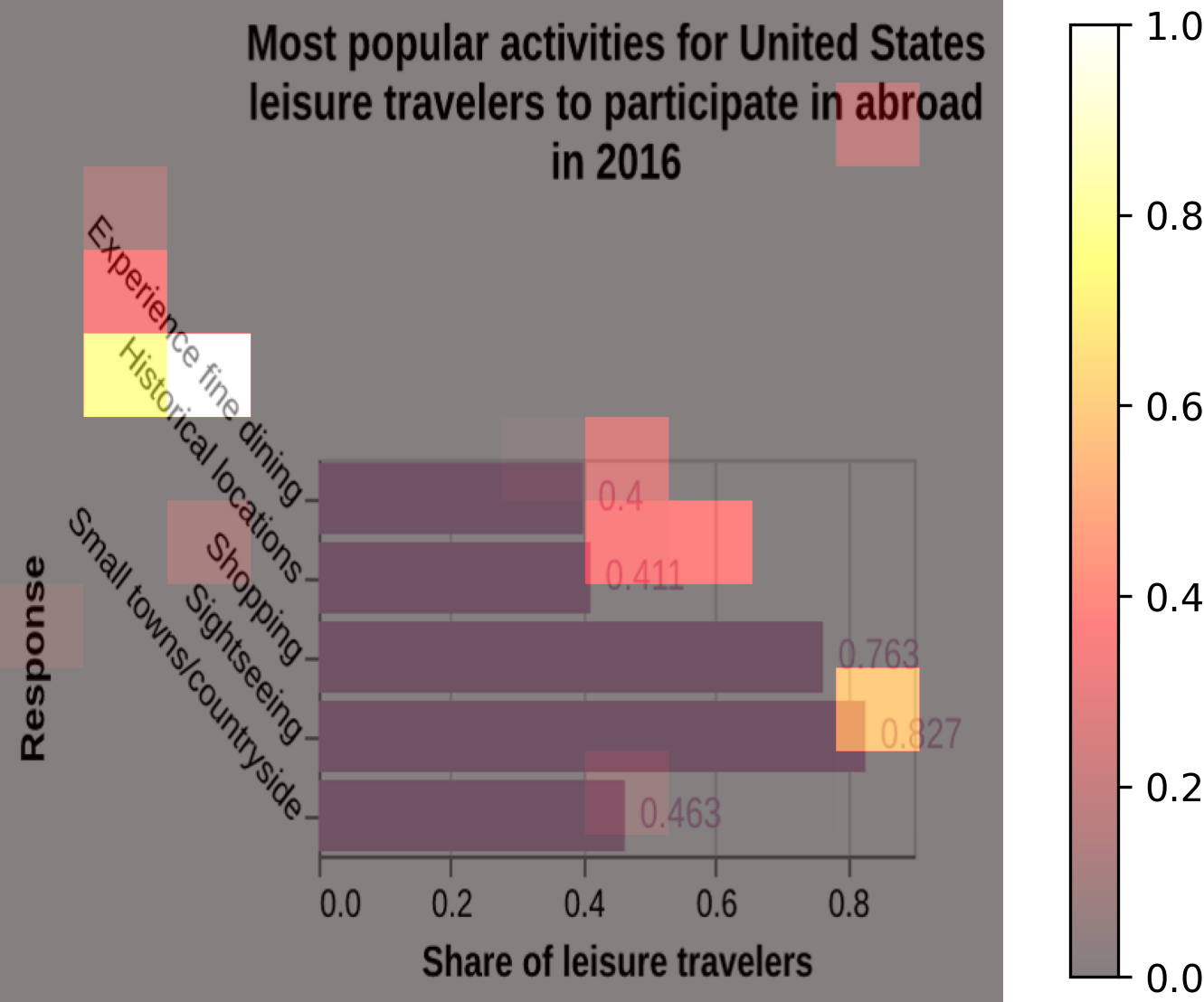}
        \caption{InternVL2-Response: \textit{``0.4"} ($\square$)}
        \label{fig:internvl-bar-anno-uncovered-correct}
    \end{subfigure}
    \hfill
    \vspace{-0.5em}
    \caption{Examples of labeled regions and importance heatmap of InternVL2 on a Bar (Anno) chart.
    Given the task \textit{``Determine the share of leisure travelers for historical locations.''}, InternVL2 incorrectly locates the bar for ``Experience fine dining," which is closely positioned near the correct one. As a result, it generates an imperfect answer, 0.4. However, as this value is within 5\% of the target value, 0.411, it is judged as correct according to the evaluation rubric considering human perception.
    }
    \vspace{-0.5em}
    \label{fig:case-study-uncovered-correct}
\end{figure*}

Figure~\ref{fig:case-study-uncovered-correct} demonstrates a case where models answer correctly even when not fully utilizing the important regions. 
This shows that, region localization abilities of LMMs diminish when information is rendered unusually, such as when categories are shown obliquely.

\section{Discussion}
In this section, we study SOTA LMMs with more complex charts from the perspective of the number of data points and the data dimensions. Both show current LMMs are not capable of handling complex charts.

\begin{figure}[tbh]
    \centering
    \includegraphics[width=0.85\linewidth]{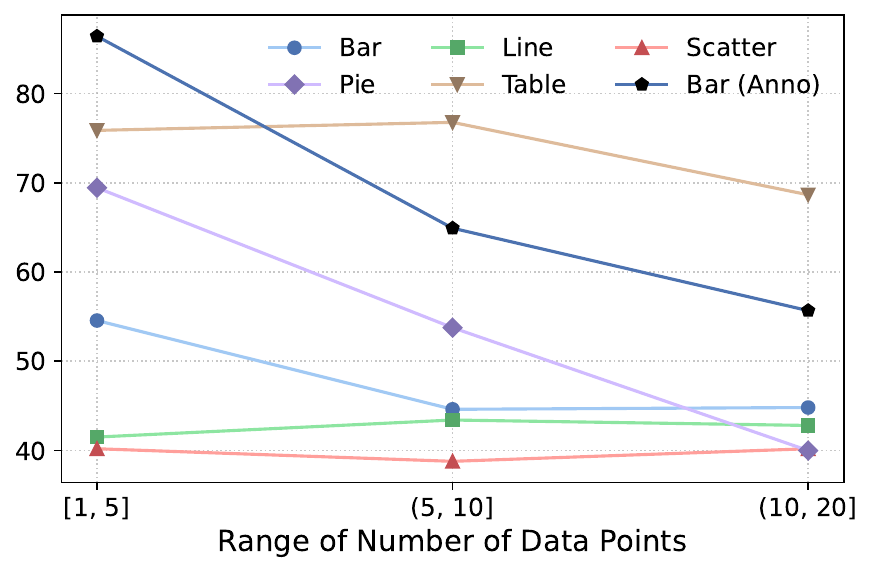}
    \caption{Overall accuracy of GPT-4o given 100 datasets with different sampled data points and different chart types.}
    \label{fig:performance-vs-number-changes}
    \vspace{-0.5em}
\end{figure}

\subsection{Performance on More Data Points}
\label{sec:data-number-change}
To measure the impact of the number of data points on LMMs' graphical perception abilities, we randomly sample 100 datasets, each containing at least 20 data points. We then systematically reduce the number of data points into three buckets: $[1, 5]$, $(5, 10]$, and $(10, 20]$, and observe how model performance varies across common chart types representing these data points.
Figure~\ref{fig:performance-vs-number-changes} shows the overall accuracy of GPT-4o when tested on 100 datasets.
As the number of data points increases, the performance consistently declines across all chart types, highlighting the model's sensitivity to data density.
Notably, Bar (Anno) charts exhibit the steepest drop in accuracy, suggesting that while numerical annotations aid graphical perception in simpler cases, the presence of more data points and numbers overwhelms the model’s ability to effectively perceive the charts.

\subsection{Performance on Multi-Dimensional Dataset}
\label{sec:multi-data-dimension}
\begin{figure}[tbh]
    \centering
    \includegraphics[width=0.85\linewidth]{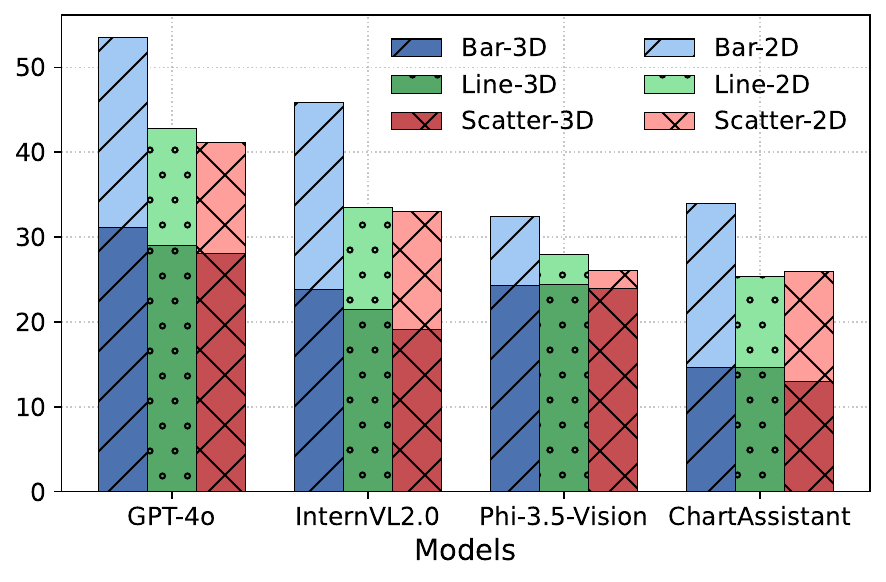}
    \caption{Overall performance comparison of four LMMs when given two-dimensional and three-dimensional datasets.}
    \label{fig:2d-3d-comparison}
    \vspace{-0.5em}
\end{figure}

We select 100 datasets from ChartLLM~\citep{Ko2024ChartLLM}, ensuring each dataset contains three data dimensions (e.g., Nominal-Numerical-Nominal) with a controlled number of data points.
These datasets are then manually edited to create the popular chart types of interest: bar, line, and scatter.
Figure~\ref{fig:2d-3d-comparison} compares the performance of various models when understanding two-dimensional (2D) and three-dimensional (3D) datasets across different chart types.
The results indicate a notable performance drop when models are tasked with three-dimensional data visualization, particularly for bar and scatter charts.
GPT-4o performs the best overall but still shows significant degradation when moving from 2D to 3D visualizations.
InternVL2 and Phi-3.5 show similar trends, though Phi-3.5 is relatively more robust than other models.
ChartAssistant performs poorly overall, with minimal adaptability between 2D and 3D contexts.
These findings indicate that current LMMs cannot fully understand advanced data visualizations yet.
Figure~\ref{fig:appendix-chart-3d} shows examples.

\section{Related Work}

\paragraph{Graphical Perception}
\citet{cleveland1984graphical} introduce graphical perception as the visual interpretation of data through basic visual elements, such as position and color.
Extensive research has since expanded on their work, evaluating human perception across diverse data types~\citep{Heer2009GPTimeSeries, Javed2010MultiTimeSeries, Whitlock2020ImmersiveAnalysis, Borkin2013ChartMemorable}, increasingly complex tasks~\citep{Saket2019TaskEffectiveness, Xiong2023ChartBias, Bearfield2024ChartSay}, and complex charts~\citep{Heer2010Crowdsourcing}. 
In the context of neural networks, prior work tests the graphical perception abilities of vision-only models~\citep{Haehn2019GraphicalPerceptionCNN} following a similar protocol.
However, evaluating and understanding the graphical perception of LMMs remains under-explored.

\paragraph{Large Multimodal Models and Their Benchmarks}
Multimodal, especially vision-and-language, modeling has evolved significantly, beginning with early models~\citep{tan2019lxmert, chen2020uniter, lu2019vilbert} that inject vision features into language understanding models, to those using contrastive learning for cross-modality representation~\citep{radford2021learning, yu2022coca, Zhang2024MagicLens}, and to the recent unified LMM frameworks~\citep{gpt-4o, openai2023gpt4v, deepmind_gemini_report, deepmind_gemini1.5_report, liu2024llava, cai2023vipllava} for various downstream tasks.
Graphical perception is widely yet implicitly considered in benchmarks for evaluating these LMMs, including task-specific ones~\citep{masry-etal-2022-chartqa, Mathew2022InfographicVQA, wang2024charxiv} and the recent holistic ones~\citep{yue2023mmmu, liu2023mmbench, lu2023mathvista, yu2023mmvet}.
Despite its ubiquity, all existing benchmarks assess graphical perception indirectly, often evaluating it alongside other abilities like reasoning or by introducing increasingly complex charts.
Another line of benchmarks~\citep{fu2024blink, tong2024cambrian} evaluates the general visual perception abilities of LMMs by repurposing traditional vision tasks, most of which focus on natural images.
These benchmarks provide limited fine-grained insights into the specific graphical perception abilities of LMMs.
Our work, instead, uniquely isolates and directly evaluates the graphical perception abilities of LMMs in a comprehensive fashion.

\section{Conclusion}

This work introduces a comprehensive and configurable evaluation framework for automatically measuring the graphical perception abilities of LMMs, offering fine-grained insights into current SOTA LMMs. 
Our findings reveal that these models struggle to generalize across diverse chart types, understand fundamental visual elements, and cross reference values within charts.
Future work may leverage this framework to synthesize diverse data for training and testing on a wider range of tasks, potentially enabling improved graphical perception and general low-level visual reasoning.
We hope the framework and these findings can help guide the development of LMMs with more generalizable perception abilities in the future.

\section{Acknowledgement}
We would like to thank our colleagues in the Deep Learning group at Microsoft Research and the OSUNLP group at the Ohio State University for their constructive feedback.
{
    \small
    \bibliographystyle{ieeenat_fullname}
    \bibliography{main}
}

\clearpage
\appendix
\setcounter{table}{0}
\setcounter{figure}{0}
\renewcommand{\thetable}{B\arabic{table}}
\renewcommand{\thefigure}{B\arabic{figure}}
\renewcommand{\theHfigure}{B\arabic{figure}}
\renewcommand{\theHtable}{B\arabic{table}}

\onecolumn
\section{Task Generation}
\label{appendix:task-generation}

\begin{tcolorbox}[colframe=green!50!black, colback=green!10!white, title=Task Generation Prompt (with text input only)]
\small 
You are a teacher to provide problems for students to solve. The problems are about understanding data and visualizations.
We will provide you with an input data, a Vega-Lite program, and a task type that the understanding task should base on. 
You will need to design a chart understanding task contextualized in the given data and chart.

\bigskip

Design the task based off one of the following idioms:

• Retrieve Value. For this task, ask students to identify values of attributes for given data points.
For example, what is the value of horsepower for Mazda CX50?

• Find Extremum: For given concrete conditions on data attribute values, ask students to find data points satisfying those conditions.
For example, which car types have the most city miles per gallon?

• Find Anomalies: ask students to identify any anomalies within a given set of data points with respect to a given relationship or expectation.
For example, which car types have abnormally low MPG?

• Determine Range: For a given set of data points and an attribute of interest, ask students to find the span of values within the set.
For example, what is the range of car prices?

• Find Correlation: for a given set of two data attributes, ask students to determine if there is a correlation between them.
For example, is there a strong correlation between car price and MPG?

• Compute Derived Value: for a given set of data points, ask students to compute an aggregate value of those data points.
For example, what is the sum of the budget for the action and the sci-fi movies?

• Filter: For given concrete conditions on data attribute values, ask students to find data points satisfying those conditions.
For example, which car types have miles per gallon ranging from 20 to 40?

• Order: For a given set of data points, ask students to rank them according to a specific ordinal metric.
For example, list the car types based on their MPG from low to high.

• Find Clusters: for a given set of data points, ask students to count the number of groups of similar data attribute values.
For example, how many different car brands are shown in the chart below?

• Characterize Distribution: for a given set of data points, ask students to identify the distribution of that attribute's values over the set.
For example, what percentage of the cars with MPG higher than 30?

\bigskip

You need to match the following requirements:
\smallskip

~~1. The task should be reasonable, and it should not exceed one sentence, and it should be contexualized in the given data.

~~2. The task should be achievable by reading the visualization without referring other tools.

~~3. The task should be self-contained with the given dataset, it should not require student to look up external information.

~~4. Each task should have a standard answer, avoid generating questions like ``compare two values of your choice."

~~5. Try not to repeat the verb for each task to maximize diversity.

\bigskip

Create a \texttt{[Task]} based off the \texttt{[Data Summary]} and \texttt{[VegaLite Script]} provided.   

The response should be in a json format:

\texttt{\{"reason":...,"tasks":[\{"description":...,"type":...\},...]\}}, including how you design the task and the actual task description. 

Generate 10 tasks at once.

\bigskip
\textbf{For example:}

\smallskip

\texttt{[Data Summary]}

\begin{verbatim}
|Date      |Location
0|5/12/2009|Houston, TX
1|4/18/2009|McAllen, TX
2|7/11/2009|Indianapolis, IN
3|11/14/2009|Kansas City, MO|MO
4|3/12/2010|Chicago, IL|IL
...
\end{verbatim}

\texttt{\{Task Demonstration\}}
% ...
\end{tcolorbox}


\section{Evaluation Rubrics}
\label{appendix:evaluation-rubric}
\begin{tcolorbox}[colframe=blue!50!black, colback=blue!10!white, title=Evaluation Prompt (\textbf{with text input only})]
\small 
You are a teacher to grade students' answers.
We will provide you a dataset, a list of tasks and student answers. Your goal is to use the dataset to evaluate if the student's answer is correct.
In order to form a good judgement, you should first use the dataset to derive your answer, and then compare it with the students asnwer.

\bigskip
When you generate the referenece answer:

* If the task asks for a value, provide the value directly.

* If the task asks for trend or correlation, answer it with one of ``increasing", ``decreasing" if the general trend point to the direcrtion, otherwise provide "unclear".

* Provide a brief reasoning of how you come up with your answer in ``reasoning" part.

* If you cannot answer a question, provide ``I don't know" as the answer, try not to provide a wrong answer.

\bigskip
When evaluating student's answer:

The student\_answer\_correctness should include the grading results of the student's answer and must be one of the following options:

\quad - correct

\quad - fair (somewhat close but not precise)

\quad - incorrect

\quad - skipped (if the student skipped the answer)

\quad - n/a (if the task does not make sense or is not answerable with the given dataset)

\bigskip
Note that if the student’s answer (value) is an approximation within 5\% of your reference answer, it is considered as correct. If is is an approximation within 20\% of your reference answer, it is fair.

For order-based tasks, such as ranking items, the student answer must match the expected orders you found.
However, for list-based tasks where order is not important, the specific sequence does not need to match as long as all relevant items are included.

\bigskip
Grade student questions based on [Data], [Tasks \& Student Answers].

The output json should have the format:

[\{``reasoning": ...,

~~~``reference answer": ...,

~~~``comparison\_with\_student\_answer": ...,

~~~``student\_answer\_correctness": ...\},

...]

\bigskip

\textbf{For example:}

\smallskip

\{Evaluation Demonstration\}
\end{tcolorbox}

We manually review GPT-4o’s and Claude’s evaluations of the same 200 questions for each of the four models.
Table~\ref{tab:evaluation-accuracy} shows that both evaluators are consistent and accurate, confirming the reliability of our findings.

%

\begin{table}[ht]
\centering
\small
\begin{tabular}{l|c|c}
\toprule
\textbf{Model} & \textbf{GPT-4o Acc. (\%)} & \textbf{Claude-3.5-Sonnet Acc. (\%)} \\ \midrule
GPT-4o         & 99.0                          & 98.5                                 \\
InternVL-2     & 99.5                          & 98.0                                 \\
Phi-3          & 99.0                          & 97.5                                 \\
ChartAssistant & 98.5                          & 98.5                                 \\ \bottomrule
\end{tabular}
\vspace{-0.5em}
\caption{Evaluation accuracy of GPT-4o and Claude-Sonnet.}
\label{tab:evaluation-accuracy}
\end{table}
\vspace{-0.5em}

\section{Model Inference}

For each model, we explored prompt design before large-scale evaluation by using default system prompt and prompts used in prior works.
We found that using different prompts does not lead to notably different results, since our charts and tasks are fairly simple and straightforward, and our findings hold across various prompts for general-purpose LMMs.
Therefore, to ensure a fair and consistent comparison, we use the same prompt (shown below) for all general-purpose LMMs and the specified prompt for ChartAssistant.
Our general prompt covers the requirements for each task type.
\begin{tcolorbox}[colframe=yellow!50!black,width=\linewidth,colback=yellow!10!white, title=Model Inference Prompt]
\footnotesize 
You are an expert in answering questions based on charts.
We will provide you with a chart and a question. Your goal is to read the chart and answer the question.

* If the question asks for a value, read the chart and provide the value directly. If you have trouble reading the exact value, provide a close estimate and indicate ``approximately".

* If the question asks for a trend, answer it with one of ``increasing", ``decreasing" if the general trend points in the direction, otherwise provide "unclear".

* Provide a brief reasoning of how you come up with your answer in "reasoning" part.
* If you cannot answer a question, provide "I don't know" as the answer, try not to provide a wrong answer.

* Answer your question based on [Chart] and [Tasks].
The output json should have the format 
\{"reasoning'': ..., ``anwer'': ...\}.
...
\end{tcolorbox}

\section{Examples of Charts}
\label{appendix:chart-examples}

\setcounter{table}{0}
\setcounter{figure}{0}
\renewcommand{\thetable}{D\arabic{table}}
\renewcommand{\thefigure}{D\arabic{figure}}
\renewcommand{\theHfigure}{D\arabic{figure}}
\renewcommand{\theHtable}{D\arabic{table}}

\begin{figure}[bt]
    \centering
    \begin{subfigure}[t]{0.31\textwidth}
        \centering
        \includegraphics[height=4.0cm]{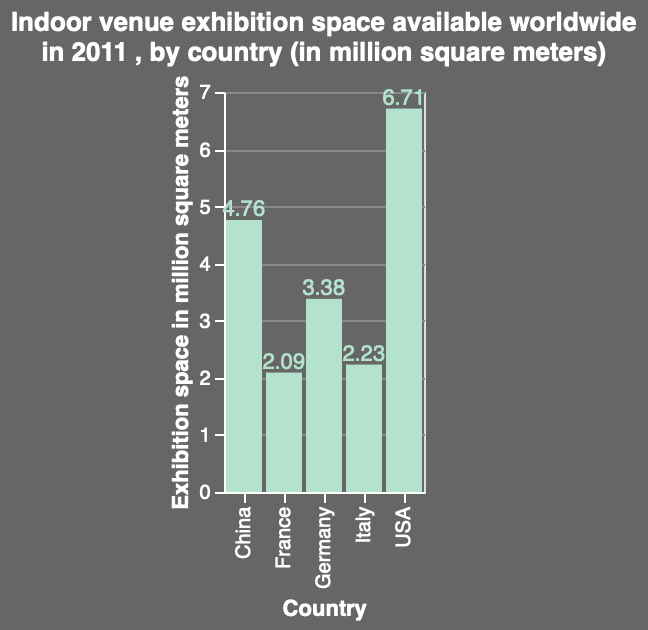}
        \caption{Bar (Anno)}
        \label{fig:768_bar_anno}
    \end{subfigure}
    \hfill
    \begin{subfigure}[t]{0.31\textwidth}
        \centering
        \includegraphics[height=4.0cm]{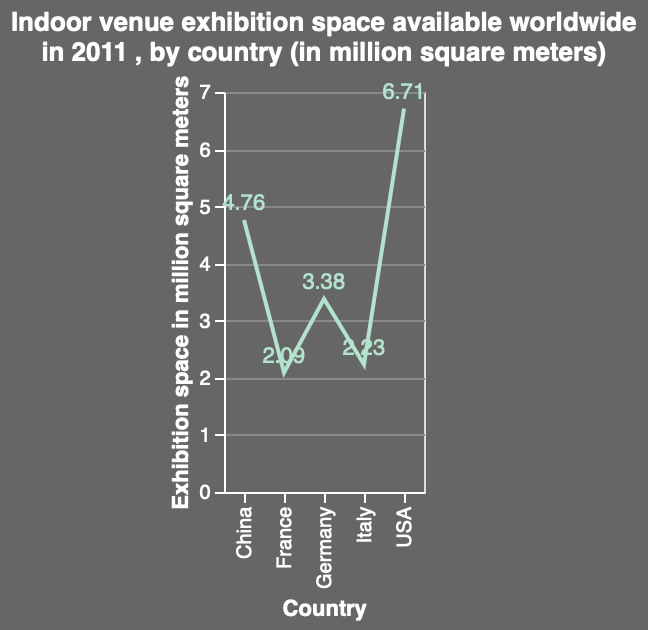}
        \caption{Line (Anno)}
        \label{fig:768_line_anno}
    \end{subfigure}
    \hfill
    \begin{subfigure}[t]{0.34\textwidth}
        \centering
        \includegraphics[height=4.0cm]{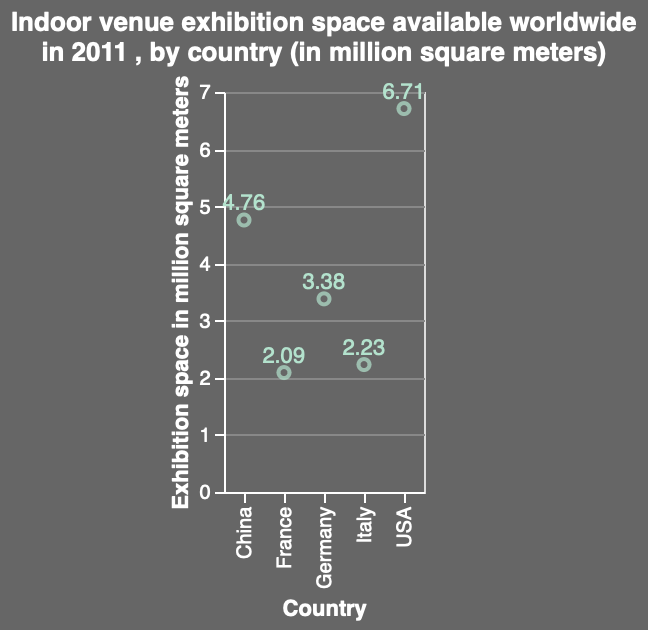}
        \caption{Scatter (Anno)}
        \label{fig:768_scatter_anno}
    \end{subfigure}

    \begin{subfigure}[t]{0.31\textwidth}
        \centering
        \includegraphics[height=4.0cm]{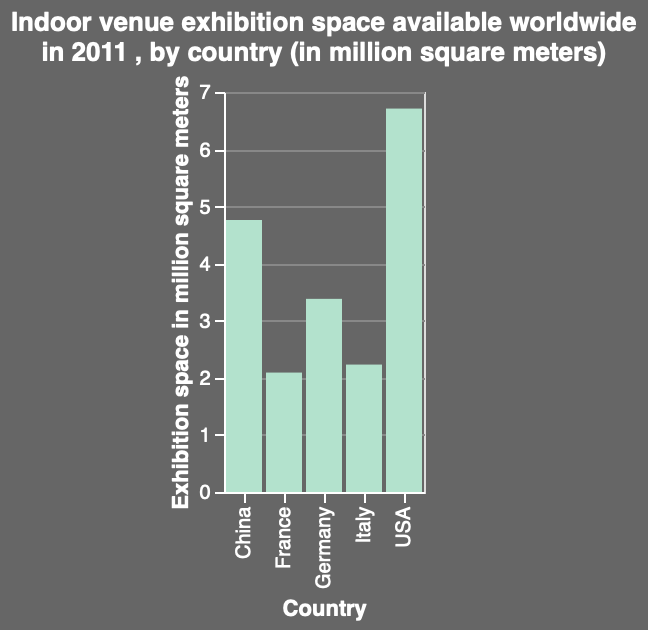}
        \caption{Bar}
        \label{fig:768_bar}
    \end{subfigure}
    \hfill
    \begin{subfigure}[t]{0.31\textwidth}
        \centering
        \includegraphics[height=4.0cm]{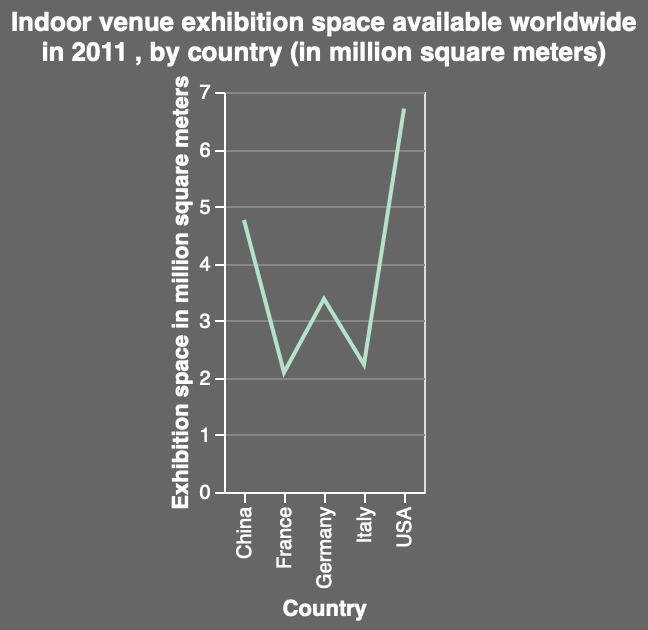}
        \caption{Line}
        \label{fig:768_line}
    \end{subfigure}
    \hfill
    \begin{subfigure}[t]{0.34\textwidth}
        \centering
        \includegraphics[height=4.0cm]{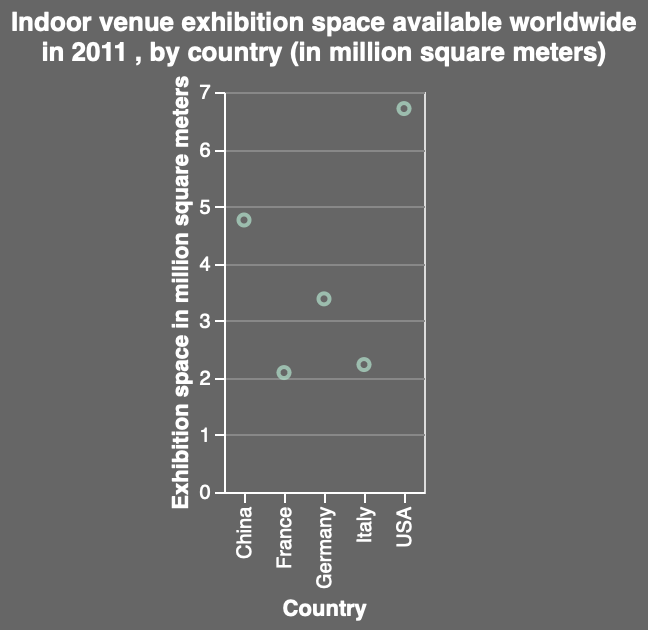}
        \caption{Scatter}
        \label{fig:768_scatter}
    \end{subfigure}

\begin{subfigure}[t]{0.48\textwidth}
    \centering
    \includegraphics[height=4.0cm]{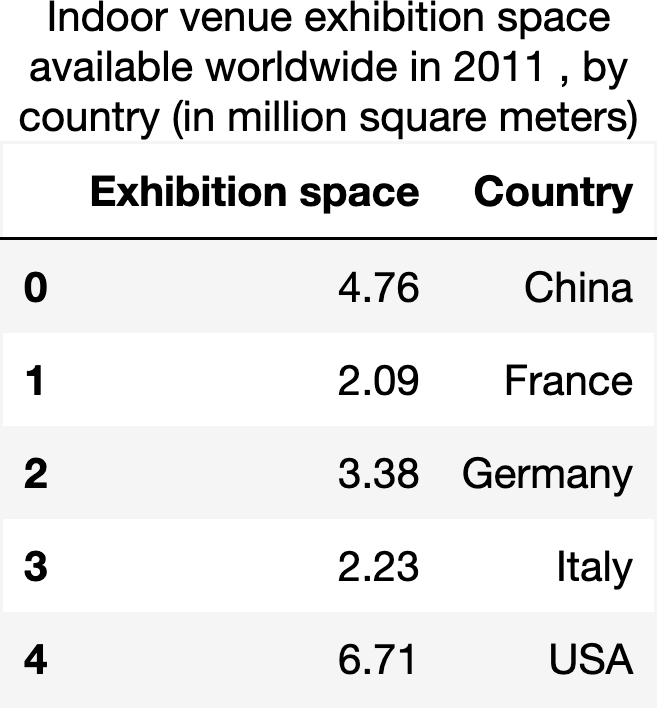}
    \caption{Table}
    \label{fig:768_table}
\end{subfigure}
\hfill
\begin{subfigure}[t]{0.48\textwidth}
    \centering
    \includegraphics[height=4.0cm]{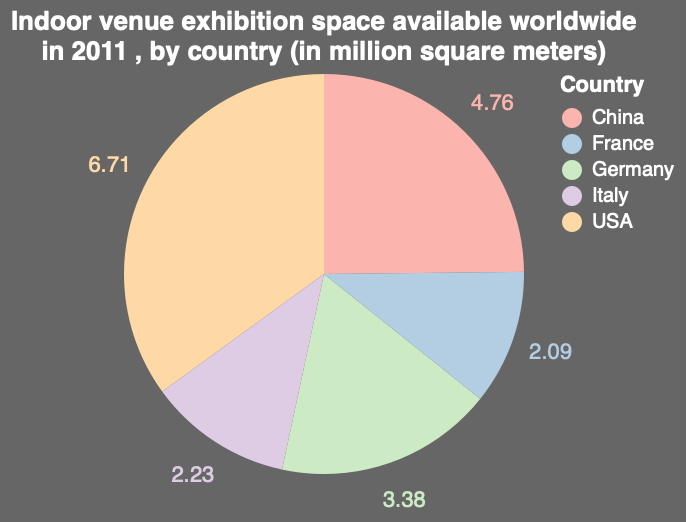}
    \caption{Pie}
    \label{fig:768_pie}
\end{subfigure}

    \caption{Cases of charts with and without numerical annotations. The Table is used as an image input for models being evaluated.
    }
    \label{fig:appendix-chart-w-and-wo-anno-cases}
\vspace{-1em}
\end{figure}

\begin{figure}[tb]
    \centering
    \begin{subfigure}[t]{0.31\textwidth}
        \centering
        \includegraphics[width=4.0cm]{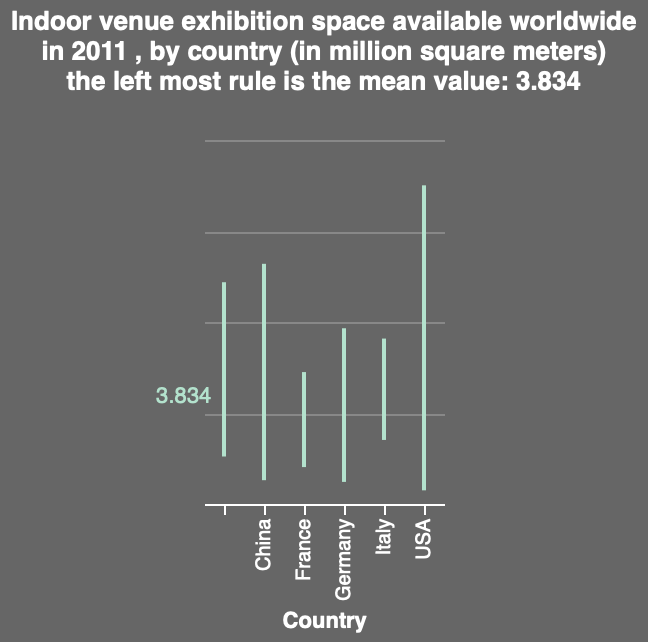}
        \caption{Length ($\leftrightarrow$)}
        \label{fig:768_length}
    \end{subfigure}
    \hfill
    \begin{subfigure}[t]{0.31\textwidth}
        \centering
        \includegraphics[width=4.0cm]{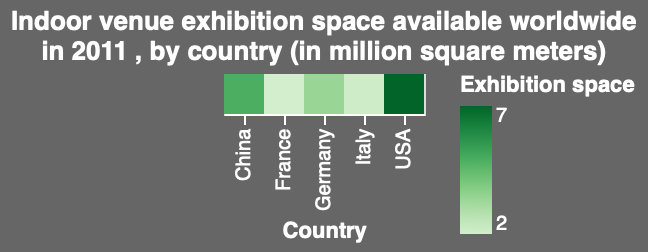}
        \caption{Color (\tikz \fill[Air Force blue] (0,0) rectangle (0.2,0.2);)}
        \label{fig:768_color}
    \end{subfigure}
    \hfill
    \begin{subfigure}[t]{0.34\textwidth}
        \centering
        \includegraphics[width=4.0cm]{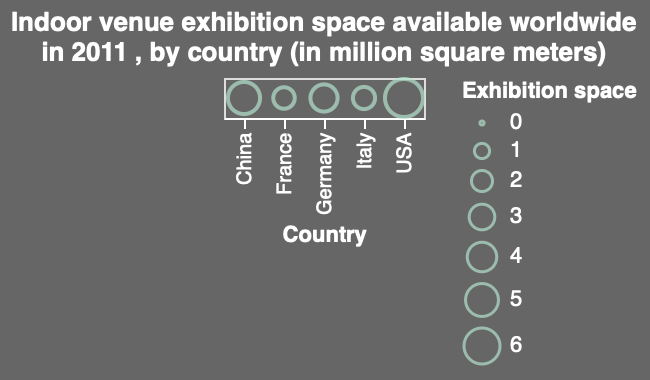}
        \caption{Size (\tikz \filldraw[fill=gray] (0,0) circle (3pt);)}
        \label{fig:768_size}
    \end{subfigure}

    \begin{subfigure}[t]{0.31\textwidth}
        \centering
        \includegraphics[width=4.0cm]{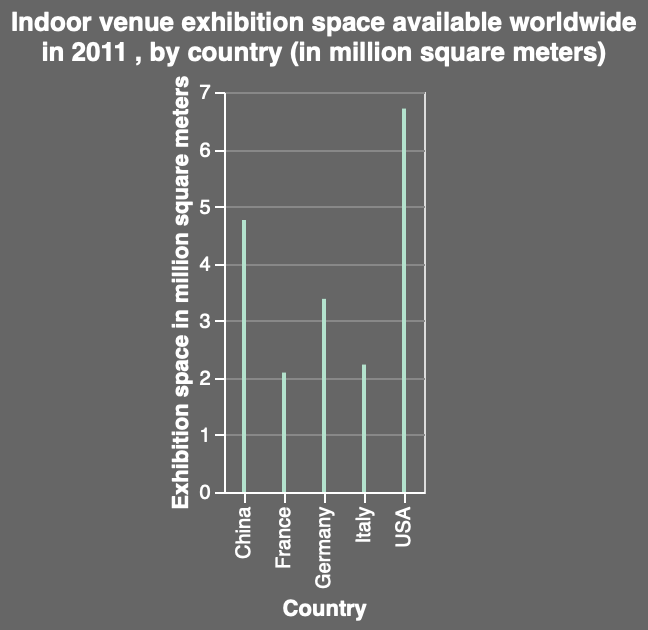}
        \caption{($\leftrightarrow$, $\star$)}
        \label{fig:768_position_length}
    \end{subfigure}
    \hfill
    \begin{subfigure}[t]{0.31\textwidth}
        \centering
        \includegraphics[width=4.0cm]{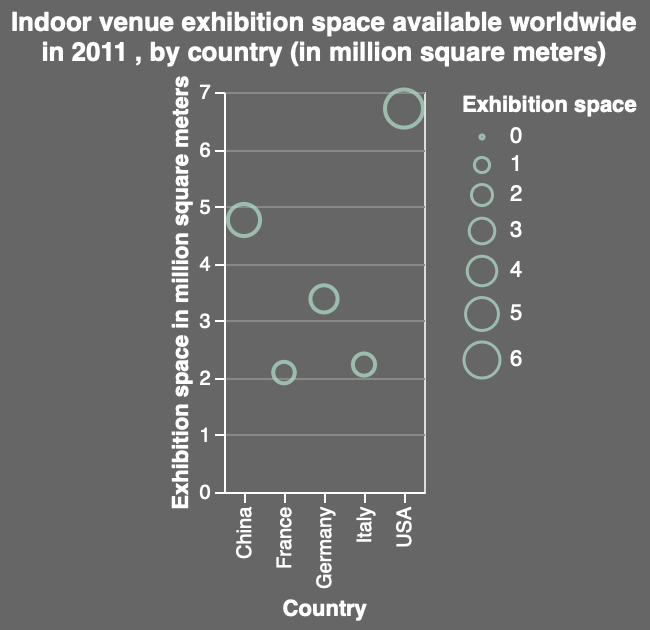}
        \caption{(\tikz \filldraw[fill=gray] (0,0) circle (3pt);, $\star$)}
        \label{fig:768_position_size}
    \end{subfigure}
    \hfill
    \begin{subfigure}[t]{0.34\textwidth}
        \centering
        \includegraphics[width=4.0cm]{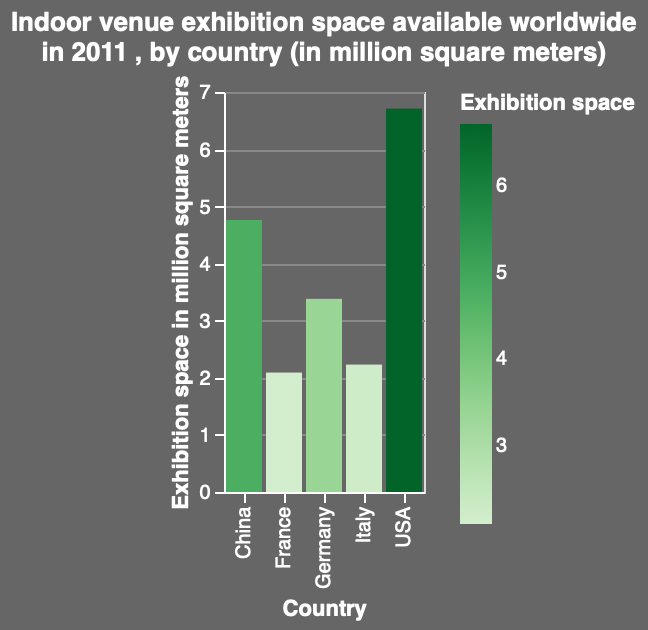}
        \caption{($\leftrightarrow$, \tikz \fill[Air Force blue] (0,0) rectangle (0.2,0.2);, \tikz \filldraw[fill=gray] (0,0) circle (3pt);, $\star$)}
        \label{fig:768_position_size_length_color}
    \end{subfigure}

    \caption{Cases of charts with single and multiple visual elements.
    }
    \label{fig:appendix-chart-w-single-and-multi-visual-elements}
\vspace{-1em}
\end{figure}

\begin{figure}[tb]
    \centering
    \begin{subfigure}[t]{0.32\textwidth}
        \centering
        \includegraphics[width=4.4cm]{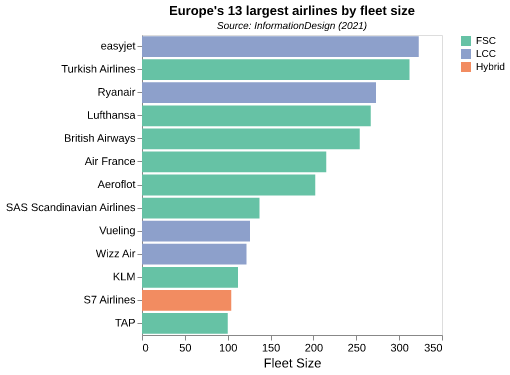}
        \caption{3D Bar-Color Hue}
        \label{fig:3d_bar}
    \end{subfigure}
    \hfill
    \begin{subfigure}[t]{0.32\textwidth}
        \centering
        \includegraphics[width=4.4cm]{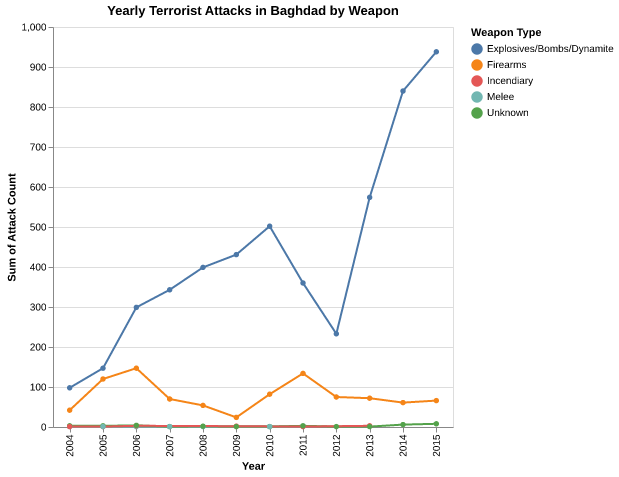}
        \caption{3D Line-Color Hue}
        \label{fig:3d_line}
    \end{subfigure}
    \hfill
    \begin{subfigure}[t]{0.32\textwidth}
        \centering
        \includegraphics[width=4.4cm]{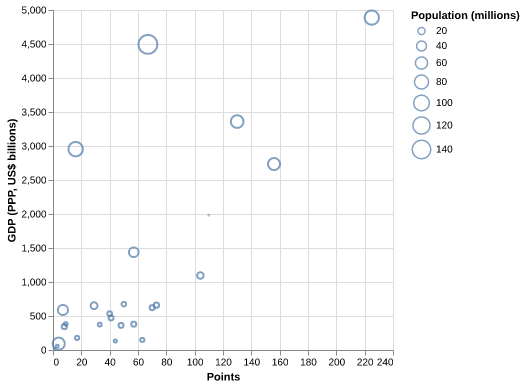}
        \caption{3D Scatter-Size}
        \label{fig:3d_scatter}
    \end{subfigure}

    \begin{subfigure}[t]{0.32\textwidth}
        \centering
        \includegraphics[width=4.4cm]{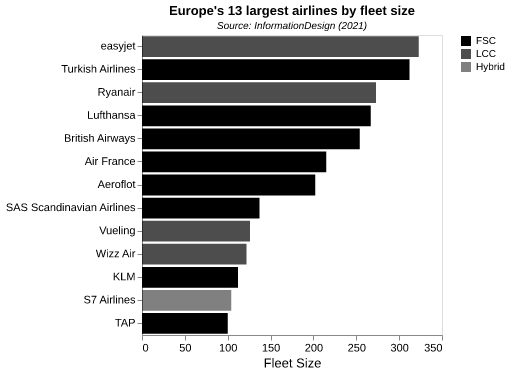}
        \caption{3D Bar-Color Luminance}
        \label{fig:3d_bar_2}
    \end{subfigure}
    \hfill
    \begin{subfigure}[t]{0.32\textwidth}
        \centering
        \includegraphics[width=4.4cm]{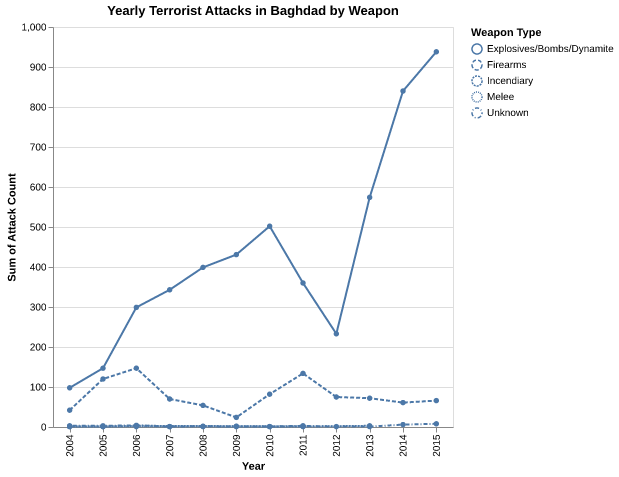}
        \caption{3D Line-Texture}
        \label{fig:3d_line_2}
    \end{subfigure}
    \hfill
    \begin{subfigure}[t]{0.32\textwidth}
        \centering
        \includegraphics[width=4.4cm]{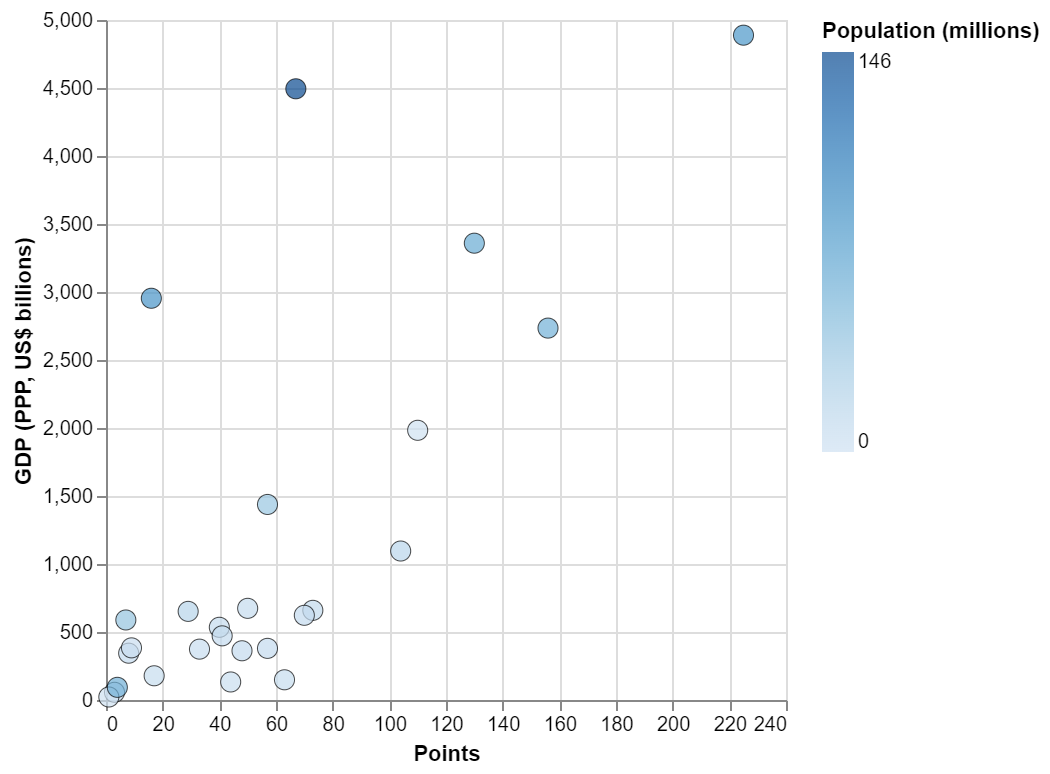}
        \caption{3D Scatter-Color Saturation}
        \label{fig:3d_scatter_2}
    \end{subfigure}
    
    \caption{Cases of charts with three data dimensions (i.e., three different data attributes).
    }
    \label{fig:appendix-chart-3d}
\vspace{-1em}
\end{figure}

We present one chart visualized in 14 different chart types used in the experiments in Figure~\ref{fig:appendix-chart-w-and-wo-anno-cases} and Figure~\ref{fig:appendix-chart-w-single-and-multi-visual-elements}.
Additionally, we include our manually edited 3D chart examples in Figure~\ref{fig:appendix-chart-3d}. These charts can also be used for direct comparisons of visual element perception, such as color hue vs. color luminance in bar charts and color hue vs. texture in line charts.
However, as current SOTA LMMs fail to achieve a satisfactory level of accuracy, we are unable to obtain meaningful insights at this time.
We leave further exploration of LMMs' visual element understanding in 3D charts for future work.


\section{Full Results}
\label{appendix:full-results}
\setcounter{table}{0}
\setcounter{figure}{0}
\renewcommand{\thetable}{E\arabic{table}}
\renewcommand{\thefigure}{E\arabic{figure}}
\renewcommand{\theHfigure}{E\arabic{figure}}
\renewcommand{\theHtable}{E\arabic{table}}
The detailed results of four representative LMMs on 10 tasks across 14 chart types used in Section~\ref{sec:rq1} and Section~\ref{sec:rq2} are shown in Tables~\ref{tab:gpt4o_all_results}, \ref{tab:internvl_all_results}, \ref{tab:phi3_all_results}, \ref{tab:chartasst_all_results}.

\begin{table}[tbh]
\centering
\resizebox{\textwidth}{!}{
\begin{tabular}{lcccccccccccccc}
\toprule
\multirow{2}{*}{} &
\multicolumn{5}{c}{w/ Number Annotated} &
\multicolumn{3}{c}{w/o Number Annotated} & \multicolumn{3}{c}{Single Element} & \multicolumn{3}{c}{Multiple Elements} \\ \cmidrule(lr){2-6} \cmidrule(lr){7-9} \cmidrule(lr){10-12} \cmidrule(lr){13-15}

& Bar & Line & Scatter & Pie & Table & Bar & Line & Scatter &
$\leftrightarrow$ &
\tikz \fill[Air Force blue] (0,0) rectangle (0.2,0.2); &
\tikz \filldraw[fill=gray] (0,0) circle (3pt); &
(\makecell{$\leftrightarrow$, $\star$}) &
(\makecell{\tikz \filldraw[fill=gray] (0,0) circle (3pt);, $\star$}) &
($\leftrightarrow$, \tikz \fill[Air Force blue] (0,0) rectangle (0.2,0.2);, \tikz \filldraw[fill=gray] (0,0) circle (3pt);, $\star$) \\
\cmidrule(lr){1-6} \cmidrule(lr){7-9} \cmidrule(lr){10-12} \cmidrule(lr){13-15}

T1 & \textbf{88.3} & 75.2 & 74.7 & 68.2 & 73.8 & 63.5 & 50.0 & 52.5 & 14.9 & 20.1 & 16.6 & 17.6 & 17.9 & 27.4 \\
T2 & \textbf{87.4} & 64.3 & 63.6 & 54.6 & 60.7 & 65.9 & 48.3 & 40.6 & 25.3 & 30.6 & 37.4 & 35.0 & 35.5 & 44.2 \\
T3 & \textbf{87.0} & 77.0 & 78.0 & 64.0 & 75.1 & 69.1 & 59.9 & 55.5 & 18.1 & 21.3 & 27.3 & 29.7 & 32.3 & 37.0 \\
T4 & 90.6 & 89.5 & 92.5 & \textbf{94.3} & 86.5 & 46.8 & 46.2 & 41.8 & 1.6 & 3.0 & 0.7 & 1.9 & 1.6 & 1.7 \\
T5 & 87.5 & 85.6 & 86.0 & 80.0 & \textbf{89.5} & 75.1 & 77.5 & 72.3 & 55.6 & 66.3 & 68.5 & 63.6 & 71.3 & 66.8 \\
T6 & 77.6 & 79.4 & \textbf{84.1} & 77.9 & 74.3 & 21.0 & 15.4 & 16.5 & 7.0 & 5.3 & 4.9 & 7.4 & 6.5 & 7.4 \\
T7 & \textbf{81.1} & 62.3 & 69.4 & 60.8 & 70.5 & 43.5 & 30.5 & 33.9 & 13.9 & 17.8 & 20.2 & 19.6 & 23.1 & 25.5 \\
T8 & \textbf{78.3} & 41.9 & 45.4 & 46.3 & 61.6 & 36.0 & 14.1 & 14.6 & 1.4 & 3.9 & 3.8 & 4.3 & 3.2 & 5.6 \\
T9 & \textbf{85.4} & 71.6 & 73.5 & 61.7 & 70.6 & 60.6 & 50.3 & 44.3 & 27.9 & 32.4 & 35.6 & 39.1 & 42.4 & 47.2 \\
T10 & \textbf{87.6} & 80.0 & 84.5 & 84.2 & 87.4 & 55.6 & 39.5 & 44.4 & 11.5 & 13.1 & 12.8 & 11.7 & 12.8 & 15.1 \\\cmidrule(lr){1-6} \cmidrule(lr){7-9} \cmidrule(lr){10-12} \cmidrule(lr){13-15}
Overall & \textbf{85.0} & 72.6 & 75.1 & 69.1 & 74.7 & 53.4 & 42.8 & 41.2 & 17.6 & 21.1 & 22.6 & 22.9 & 24.4 & 27.7 \\

\bottomrule
\end{tabular}
}
\caption{All results of GPT-4o~\citep{gpt-4o} on 14 types of charts across 10 task types. The best result on each task is marked in bold.}
\label{tab:gpt4o_all_results}

\end{table}
\begin{table}[tbh]
\centering
\resizebox{\textwidth}{!}{
\begin{tabular}{lcccccccccccccc}
\toprule
\multirow{2}{*}{} &
\multicolumn{5}{c}{w/ Number Annotated} &
\multicolumn{3}{c}{w/o Number Annotated} & \multicolumn{3}{c}{Single Element} & \multicolumn{3}{c}{Multiple Elements} \\ \cmidrule(lr){2-6} \cmidrule(lr){7-9} \cmidrule(lr){10-12} \cmidrule(lr){13-15}

& Bar & Line & Scatter & Pie & Table & Bar & Line & Scatter &
$\leftrightarrow$ &
\tikz \fill[Air Force blue] (0,0) rectangle (0.2,0.2); &
\tikz \filldraw[fill=gray] (0,0) circle (3pt); &
(\makecell{$\leftrightarrow$, $\star$}) &
(\makecell{\tikz \filldraw[fill=gray] (0,0) circle (3pt);, $\star$}) &
($\leftrightarrow$, \tikz \fill[Air Force blue] (0,0) rectangle (0.2,0.2);, \tikz \filldraw[fill=gray] (0,0) circle (3pt);, $\star$) \\
\cmidrule(lr){1-6} \cmidrule(lr){7-9} \cmidrule(lr){10-12} \cmidrule(lr){13-15}

T1 & 72.3 & 65.4 & 66.1 & 61.9 & \textbf{84.5} & 51.9 & 38.9 & 40.5 & 33.7 & 40.9 & 37.3 & 33.7 & 40.9 & 40.5 \\
T2 & 77.3 & 50.8 & 53.0 & 49.5 & \textbf{92.0} & 75.3 & 49.7 & 46.5 & 26.1 & 23.3 & 30.4 & 26.1 & 23.3 & 46.5 \\
T3 & 51.3 & 44.3 & 44.5 & 32.7 & 51.8 & \textbf{52.4} & 43.5 & 39.3 & 11.7 & 15.5 & 18.8 & 11.7 & 15.5 & 39.3 \\
T4 & 69.7 & 63.8 & 66.6 & 51.4 & \textbf{75.4} & 27.3 & 26.0 & 21.5 & 5.7 & 16.1 & 6.2 & 5.7 & 16.1 & 21.5 \\
T5 & 49.6 & 45.0 & 53.5 & 37.8 & \textbf{53.2} & 42.6 & 45.0 & 45.5 & 41.6 & 42.0 & 42.0 & 41.6 & 42.0 & 45.5 \\
T6 & 72.1 & 64.8 & 68.5 & 59.8 & \textbf{73.5} & 52.6 & 37.4 & 40.8 & 23.6 & 23.0 & 23.7 & 23.6 & 23.0 & 40.8 \\
T7 & 40.9 & 30.4 & 31.2 & 25.4 & \textbf{50.8} & 35.8 & 20.6 & 24.0 & 8.1 & 10.1 & 12.6 & 8.1 & 10.1 & 24.0 \\
T8 & 58.2 & 27.8 & 32.6 & 47.0 & \textbf{66.4} & 50.4 & 15.4 & 16.5 & 4.3 & 5.8 & 6.3 & 4.3 & 5.8 & 16.5 \\
T9 & 34.9 & 32.2 & 34.7 & 26.0 & \textbf{39.0} & 34.8 & 27.3 & 30.4 & 14.4 & 17.3 & 20.9 & 14.4 & 17.3 & 30.4 \\
T10 & 43.6 & 35.6 & 35.8 & 29.4 & \textbf{49.7} & 29.7 & 22.3 & 26.6 & 12.8 & 13.1 & 11.9 & 12.8 & 13.1 & 26.7 \\ \cmidrule(lr){1-6} \cmidrule(lr){7-9} \cmidrule(lr){10-12} \cmidrule(lr){13-15}
Overall & 57.6 & 46.4 & 49.0 & 42.5 & \textbf{64.2} & 45.9 & 33.0 & 33.4 & 18.3 & 20.7 & 21.1 & 25.2 & 24.1 & 26.5 \\

\bottomrule
\end{tabular}
}
\caption{All results of InternVL2~\citep{chen2024internvl2} on 14 types of charts across 10 task types. The best result on each task is marked in bold.}
\label{tab:internvl_all_results}

\end{table}
\begin{table}[tbh]
\centering
\resizebox{\textwidth}{!}{
\begin{tabular}{lcccccccccccccc}
\toprule
\multirow{2}{*}{} &
\multicolumn{5}{c}{w/ Number Annotated} &
\multicolumn{3}{c}{w/o Number Annotated} & \multicolumn{3}{c}{Single Element} & \multicolumn{3}{c}{Multiple Elements} \\ \cmidrule(lr){2-6} \cmidrule(lr){7-9} \cmidrule(lr){10-12} \cmidrule(lr){13-15}

& Bar & Line & Scatter & Pie & Table & Bar & Line & Scatter &
$\leftrightarrow$ &
\tikz \fill[Air Force blue] (0,0) rectangle (0.2,0.2); &
\tikz \filldraw[fill=gray] (0,0) circle (3pt); &
(\makecell{$\leftrightarrow$, $\star$}) &
(\makecell{\tikz \filldraw[fill=gray] (0,0) circle (3pt);, $\star$}) &
($\leftrightarrow$, \tikz \fill[Air Force blue] (0,0) rectangle (0.2,0.2);, \tikz \filldraw[fill=gray] (0,0) circle (3pt);, $\star$) \\
\cmidrule(lr){1-6} \cmidrule(lr){7-9} \cmidrule(lr){10-12} \cmidrule(lr){13-15}

T1 & 70.5 & 62.7 & 63.2 & 53.6 & \textbf{85.9} & 36.3 & 26.8 & 34.7 & 32.2 & 36.7 & 32.1 & 32.0 & 37.5 & 47.0 \\
T2 & 72.7 & 53.6 & 52.9 & 30.7 & \textbf{90.3} & 59.7 & 38.5 & 40.6 & 21.0 & 22.6 & 26.1 & 31.2 & 28.5 & 33.6 \\
T3 & 39.7 & 38.2 & 36.8 & 24.3 & \textbf{47.4} & 37.6 & 38.9 & 30.6 & 10.8 & 11.0 & 13.7 & 17.8 & 12.3 & 15.0 \\
T4 & \textbf{70.6} & 67.4 & 66.5 & 51.2 & 54.8 & 14.6 & 13.1 & 14.3 & 8.9 & 15.3 & 9.0 & 11.7 & 9.6 & 10.8 \\
T5 & 62.5 & 60.0 & \textbf{64.4} & 42.0 & 62.3 & 54.5 & 54.6 & 57.3 & 53.9 & 48.7 & 50.9 & 55.3 & 50.7 & 53.5 \\
T6 & 66.8 & 62.4 & 66.1 & 53.5 & \textbf{76.4} & 27.6 & 18.7 & 27.1 & 19.2 & 17.5 & 17.8 & 17.4 & 19.8 & 21.3 \\
T7 & 37.2 & 27.5 & 28.5 & 19.3 & \textbf{50.6} & 19.5 & 15.7 & 17.5 & 5.9 & 7.4 & 9.6 & 10.3 & 10.7 & 10.7 \\
T8 & 40.7 & 24.1 & 27.8 & 21.6 & \textbf{73.7} & 27.5 & 10.3 & 15.1 & 6.6 & 7.6 & 9.3 & 7.4 & 9.9 & 8.0 \\
T9 & 24.5 & 21.3 & 22.2 & 14.3 & \textbf{27.1} & 21.5 & 25.6 & 21.1 & 10.8 & 10.7 & 14.4 & 15.0 & 15.3 & 13.3 \\
T10 & 32.1 & 26.3 & 29.8 & 14.5 & \textbf{41.5} & 21.6 & 17.2 & 19.7 & 9.6 & 12.2 & 13.8 & 13.7 & 15.1 & 14.2 \\ \cmidrule(lr){1-6} \cmidrule(lr){7-9} \cmidrule(lr){10-12} \cmidrule(lr){13-15}
Overall & 52.2 & 44.7 & 46.2 & 32.7 & \textbf{61.6} & 32.4 & 26.0 & 27.9 & 17.8 & 18.9 & 19.6 & 21.2 & 20.9 & 22.7 \\

\bottomrule
\end{tabular}
}
\caption{All results of Phi-3.5~\citep{Abdin2024phi3} on 14 types of charts across 10 task types. The best result on each task is marked in bold.}
\label{tab:phi3_all_results}

\end{table}
\begin{table}[tbh]
\centering
\resizebox{\textwidth}{!}{
\begin{tabular}{lcccccccccccccc}
\toprule
\multirow{2}{*}{} &
\multicolumn{5}{c}{w/ Number Annotated} &
\multicolumn{3}{c}{w/o Number Annotated} & \multicolumn{3}{c}{Single Element} & \multicolumn{3}{c}{Multiple Elements} \\ \cmidrule(lr){2-6} \cmidrule(lr){7-9} \cmidrule(lr){10-12} \cmidrule(lr){13-15}

& Bar & Line & Scatter & Pie & Table & Bar & Line & Scatter &
$\leftrightarrow$ &
\tikz \fill[Air Force blue] (0,0) rectangle (0.2,0.2); &
\tikz \filldraw[fill=gray] (0,0) circle (3pt); &
(\makecell{$\leftrightarrow$, $\star$}) &
(\makecell{\tikz \filldraw[fill=gray] (0,0) circle (3pt);, $\star$}) &
($\leftrightarrow$, \tikz \fill[Air Force blue] (0,0) rectangle (0.2,0.2);, \tikz \filldraw[fill=gray] (0,0) circle (3pt);, $\star$) \\
\cmidrule(lr){1-6} \cmidrule(lr){7-9} \cmidrule(lr){10-12} \cmidrule(lr){13-15}

T1 & 54.4 & 42.5 & 43.6 & 32.8 & \textbf{55.0} & 32.8 & 23.1 & 23.6 & 9.1 & 17.9 & 14.4 & 37.8 & 28.3 & 37.3 \\
T2 & \textbf{70.2} & 46.1 & 43.2 & 28.0 & 42.9 & 70.1 & 42.6 & 44.4 & 22.9 & 24.0 & 25.2 & 34.7 & 30.2 & 35.8 \\
T3 & 47.9 & 35.7 & 34.3 & 28.7 & 34.0 & \textbf{49.9} & 38.7 & 38.5 & 12.2 & 12.1 & 12.1 & 19.6 & 16.3 & 16.4 \\
T4 & \textbf{25.7} & 20.7 & 22.3 & 11.1 & 21.6 & 8.2 & 6.6 & 6.9 & 0.7 & 2.5 & 1.5 & 2.6 & 1.3 & 1.9 \\
T5 & 54.6 & 54.9 & 49.3 & 46.9 & 44.1 & 56.2 & \textbf{58.5} & 52.7 & 41.6 & 47.7 & 45.0 & 46.2 & 49.3 & 44.8 \\
T6 & 28.1 & 21.5 & 20.9 & 18.1 & 15.7 & \textbf{32.1} & 24.0 & 25.5 & 7.3 & 7.0 & 9.3 & 13.2 & 12.6 & 13.3 \\
T7 & 22.9 & 17.0 & 16.8 & 12.1 & 17.2 & \textbf{23.5} & 18.7 & 16.6 & 8.8 & 9.3 & 9.7 & 14.7 & 11.9 & 12.0 \\
T8 & 14.7 & 5.9 & 5.2 & 4.1 & \textbf{29.6} & 14.4 & 3.7 & 3.9 & 0.8 & 1.1 & 2.6 & 1.4 & 2.2 & 1.0 \\
T9 & \textbf{29.2} & 25.2 & 22.2 & 19.4 & 15.5 & 28.7 & 27.1 & 25.0 & 11.6 & 9.7 & 8.4 & 12.8 & 11.9 & 13.6 \\
T10 & 14.5 & 13.8 & 12.2 & 10.6 & 10.0 & \textbf{17.5} & 13.5 & 14.2 & 8.2 & 9.3 & 7.2 & 9.2 & 10.8 & 8.7 \\ \cmidrule(lr){1-6} \cmidrule(lr){7-9} \cmidrule(lr){10-12} \cmidrule(lr){13-15}
Overall & \textbf{36.6} & 28.4 & 27.0 & 21.2 & 28.5 & 33.9 & 25.9 & 25.4 & 12.3 & 14.0 & 13.6 & 19.2 & 17.4 & 18.5 \\

\bottomrule
\end{tabular}
}
\caption{All results of ChartAssistant~\citep{meng2024chartassisstant} on 14 types of charts across 10 task types. The best result on each task is marked in bold.}
\label{tab:chartasst_all_results}

\end{table}

\newpage
\section{More Results on Latest Models and Human Performance}

Our evaluation framework allows us to easily incorporate more recent models to assess their graphical perception abilities. We extend our evaluation with Claude-3.5-Sonnet~\cite{Claude3}, Gemini-1.5-Pro~\cite{deepmind_gemini1.5_report}, and Qwen2-VL-72B~\cite{Wang2024Qwen2VL}.
As shown in Table~\ref{tab:rebuttal_results}, our main findings still hold for these models.
They remain unable to generalize well across chart types or effectively understand fundamental visual elements that humans can easily do.

In terms of human performance, we invite 27 college-level crowdsourced workers to solve the subset tasks used in our paper.
Each worker is assigned one chart type and random tasks, ensuring no two charts from the same dataset are seen to prevent answer leakage.
The results in Table~\ref{tab:rebuttal_results} show a significant gap between the latest models and human performance.
More importantly, humans perform consistently across different chart types, while models exhibit strong variability, particularly struggling with less common chart types.

\begin{table}[tbh]
\centering
\resizebox{0.96\textwidth}{!}{
\begin{tabular}{lcccccccccccccc}
\toprule
\multirow{2}{*}{} &
\multicolumn{5}{c}{w/ Number Annotated} &
\multicolumn{3}{c}{w/o Number Annotated} & \multicolumn{3}{c}{Single Element} & \multicolumn{3}{c}{Multiple Elements} \\ \cmidrule(lr){2-6} \cmidrule(lr){7-9} \cmidrule(lr){10-12} \cmidrule(lr){13-15}

& Bar & Line & Scatter & Pie & Table & Bar & Line & Scatter &
$\leftrightarrow$ &
\tikz \fill[Air Force blue] (0,0) rectangle (0.2,0.2); &
\tikz \filldraw[fill=gray] (0,0) circle (3pt); &
(\makecell{$\leftrightarrow$, $\star$}) &
(\makecell{\tikz \filldraw[fill=gray] (0,0) circle (3pt);, $\star$}) &
($\leftrightarrow$, \tikz \fill[Air Force blue] (0,0) rectangle (0.2,0.2);, \tikz \filldraw[fill=gray] (0,0) circle (3pt);, $\star$) \\
\cmidrule(lr){1-6} \cmidrule(lr){7-9} \cmidrule(lr){10-12} \cmidrule(lr){13-15}

Claude-3.5-Sonnet & 85.7 & 78.8 & 77.5 & 80.2 & \textbf{95.2} & 68.7 & 55.1 & 59.6 & 20.8 & 23.7 & 23.1 & 33.2 & 28.0 & 34.4 \\
Gemini-1.5-Pro & 77.5 & 69.1 & 71.2 & 62.5 & \textbf{91.0} & 60.6 & 45.1 & 55.1 & 16.5 & 18.4 & 19.2 & 31.2	 & 26.0 & 27.8 \\
Qwen2-VL-72B  & 68.0 & 59.2 &  60.0 & 55.8 & \textbf{81.2} & 59.1 & 41.3 & 46.3 & 18.2 & 21.4 & 22.0 & 29.1	& 26.9 & 29.8 \\
\cmidrule(lr){1-6} \cmidrule(lr){7-9} \cmidrule(lr){10-12} \cmidrule(lr){13-15} 

Human (1/10 Subset)  & \textbf{98.4} & 97.8 & 97.6 & 89.1 & 97.4 & 96.2 & 95.3 & 94.8 & 82.3 & 82.7 & 81.4 & 95.6 & 94.5 & 96.5 \\

\bottomrule
\end{tabular}
}
\vspace{-0.5em}
\caption{Average performance across 10 task types on latest models.}
\label{tab:rebuttal_results}
\end{table}

\end{document}